\input harvmac
%\draftmode
\let\includefigures=\iftrue
\let\useblackboard=\iftrue
\newfam\black

%Figure Stuff
\includefigures
\message{If you do not have epsf.tex (to include figures),}
\message{change the option at the top of the tex file.}
\input epsf
\def\figin{\epsfcheck\figin}\def\figins{\epsfcheck\figins}
\def\epsfcheck{\ifx\epsfbox\UnDeFiNeD
\message{(NO epsf.tex, FIGURES WILL BE IGNORED)}
\gdef\figin##1{\vskip2in}\gdef\figins##1{\hskip.5in}% blank space instead
\else\message{(FIGURES WILL BE INCLUDED)}%
\gdef\figin##1{##1}\gdef\figins##1{##1}\fi}
\def\DefWarn#1{}
\def\figinsert{\goodbreak\midinsert}
\def\ifig#1#2#3{\DefWarn#1\xdef#1{fig.~\the\figno}
\writedef{#1\leftbracket fig.\noexpand~\the\figno}%
\figinsert\figin{\centerline{#3}}\medskip\centerline{\vbox{
\baselineskip12pt\advance\hsize by -1truein
\noindent\footnotefont{\bf Fig.~\the\figno:} #2}}
%\bigskip
\endinsert\global\advance\figno by1}
%%%
\else
\def\ifig#1#2#3{\xdef#1{fig.~\the\figno}
\writedef{#1\leftbracket fig.\noexpand~\the\figno}%
%\figinsert\figin{\centerline{#3}}\medskip
%\centerline{\vbox{\baselineskip12pt
%\advance\hsize by -1truein\noindent
%\footnotefont{\bf Fig.~\the\figno:} #2}}
%\bigskip\endinsert
\global\advance\figno by1} \fi

\def\id{{1 \kern-.28em {\rm l}}}

\def\CM{{\cal M}}

\def\K3{{\bf K3}}
\def\journal#1&#2(#3){\unskip, \sl #1\ \bf #2 \rm(19#3) }
\def\andjournal#1&#2(#3){\sl #1~\bf #2 \rm (19#3) }

\def\bar{\overline}

\def\ie{{\it i.e.}}
\def\eg{{\it e.g.}}

\def\tilde{\widetilde}

\def\frac#1#2{{#1\over#2}}

\def\inbar{\,\vrule height1.5ex width.4pt depth0pt}
\def\IC{\relax\hbox{$\inbar\kern-.3em{\rm C}$}}
\def\IR{\relax{\rm I\kern-.18em R}}
\def\IP{\relax{\rm I\kern-.18em P}}

%
%%%%%%%%%%%%%%%%%%%%%%%%%%%%%%%%%%%%
%

\def\cmp#1#2#3{Comm. Math. Phys. {\bf #1} (#2) #3}

\catcode`\@=11
\def\slash#1{\mathord{\mathpalette\c@ncel{#1}}}
\overfullrule=0pt

\def\CC{{\cal C}}

\def\MM{{\cal M}}

\def\underrel#1\over#2{\mathrel{\mathop{\kern\z@#1}\limits_{#2}}}

\catcode`\@=12

%%%%%%%%%%%%%%%%%%%%%%%%%%%%%%%%%%%%%%%%%%%%%%%%%%%%%%%%%%%%%%

%

\def\det{{\rm det}}

\def \cosh{{\rm cosh}}

\def\det{{\rm det}}
\def\exp{{\rm exp}}

%%%%%%%%%%%%%%%%%%%%%%%%%%%%%%%%%%%%%%%%%%%%%%%%%%%%%%%%%%%%%%
% new defs:

\def\ie{{\it i.e.}}
\def\eg{{\it e.g.}}

%%%%%%%%%%%%%%%%%%%%%%%%%%%%%%%%%%%%%%%%%%%%%%%%%%
%\KlebanovHB
\lref\KlebanovHB{
  I.~R.~Klebanov and M.~J.~Strassler,
  ``Supergravity and a confining gauge theory: Duality cascades and
  chiSB-resolution of naked singularities,''
  JHEP {\bf 0008}, 052 (2000)
  [arXiv:hep-th/0007191].
  %%CITATION = JHEPA,0008,052;%%
}

%\DymarskyXT
\lref\DymarskyXT{
  A.~Dymarsky, I.~R.~Klebanov and N.~Seiberg,
  ``On the moduli space of the cascading SU(M+p) x SU(p) gauge theory,''
  JHEP {\bf 0601}, 155 (2006)
  [arXiv:hep-th/0511254].
  %%CITATION = JHEPA,0601,155;%%
}

%\GiveonSR
\lref\GiveonSR{
  A.~Giveon and D.~Kutasov,
  ``Brane dynamics and gauge theory,''
  Rev.\ Mod.\ Phys.\  {\bf 71}, 983 (1999)
  [arXiv:hep-th/9802067].
  %%CITATION = RMPHA,71,983;%%
}

%\AharonyMI
\lref\AharonyMI{
  O.~Aharony, D.~Kutasov, O.~Lunin, J.~Sonnenschein and S.~Yankielowicz,
  ``Holographic MQCD,''
  Phys.\ Rev.\  D {\bf 82}, 106006 (2010)
  [arXiv:1006.5806 [hep-th]].
  %%CITATION = PHRVA,D82,106006;%%
}

%\ElitzurFH
\lref\ElitzurFH{
  S.~Elitzur, A.~Giveon and D.~Kutasov,
  ``Branes and N = 1 duality in string theory,''
  Phys.\ Lett.\  B {\bf 400}, 269 (1997)
  [arXiv:hep-th/9702014].
  %%CITATION = PHLTA,B400,269;%%
}

%\McOristIN
\lref\McOristIN{
  J.~McOrist and A.~B.~Royston,
  ``Relating Conifold Geometries to NS5-branes,''
  arXiv:1101.3552 [hep-th].
  %%CITATION = ARXIV:1101.3552;%%
}

%\GiveonFK
\lref\GiveonFK{
  A.~Giveon and D.~Kutasov,
  ``Gauge symmetry and supersymmetry breaking from intersecting branes,''
  Nucl.\ Phys.\  B {\bf 778}, 129 (2007)
  [arXiv:hep-th/0703135].
  %%CITATION = NUPHA,B778,129;%%
}

%\GiveonEW
\lref\GiveonEW{
  A.~Giveon and D.~Kutasov,
  ``Stable and Metastable Vacua in Brane Constructions of SQCD,''
  JHEP {\bf 0802}, 038 (2008)
  [arXiv:0710.1833 [hep-th]].
  %%CITATION = JHEPA,0802,038;%%
}

%\GiveonEF
\lref\GiveonEF{
  A.~Giveon and D.~Kutasov,
  ``Stable and Metastable Vacua in SQCD,''
Nucl.\ Phys.\ B {\bf 796}, 25 (2008).
[arXiv:0710.0894 [hep-th]].
%%CITATION = arXiv:0710.0894%%
}

%\GiveonUR
\lref\GiveonUR{
  A.~Giveon, D.~Kutasov, J.~McOrist and A.~B.~Royston,
  ``D-Terms and Supersymmetry Breaking from Branes,''
  Nucl.\ Phys.\  B {\bf 822}, 106 (2009)
  [arXiv:0904.0459 [hep-th]].
  %%CITATION = NUPHA,B822,106;%%
}

%\SeibergPQ
\lref\SeibergPQ{
  N.~Seiberg,
  ``Electric - magnetic duality in supersymmetric nonAbelian gauge theories,''
Nucl.\ Phys.\ B {\bf 435}, 129 (1995).
[hep-th/9411149].
%%CITATION = hep-th/9411149%%
}

%\WittenEP
\lref\WittenEP{
  E.~Witten,
  ``Branes and the dynamics of {QCD},''
  Nucl.\ Phys.\  B {\bf 507}, 658 (1997)
  [arXiv:hep-th/9706109].
  %%CITATION = NUPHA,B507,658;%%
}

%\StrasslerQS
\lref\StrasslerQS{
  M.~J.~Strassler,
  ``The duality cascade,''
  arXiv:hep-th/0505153.
  %%CITATION = HEP-TH/0505153;%%
}

%\IntriligatorDD
\lref\IntriligatorDD{
  K.~A.~Intriligator, N.~Seiberg and D.~Shih,
  ``Dynamical SUSY breaking in meta-stable vacua,''
  JHEP {\bf 0604}, 021 (2006)
  [arXiv:hep-th/0602239].
  %%CITATION = JHEPA,0604,021;%%
}

%\GiveonWP
\lref\GiveonWP{
  A.~Giveon, A.~Katz and Z.~Komargodski,
  ``On SQCD with massive and massless flavors,''
  JHEP {\bf 0806}, 003 (2008)
  [arXiv:0804.1805 [hep-th]].
  %%CITATION = JHEPA,0806,003;%%
}

\lref\gkunpub{A. Giveon and D. Kutasov, unpublished.}

%\AbelCR
\lref\AbelCR{
  S.~A.~Abel, C.~S.~Chu, J.~Jaeckel and V.~V.~Khoze,
  ``SUSY breaking by a metastable ground state: Why the early universe
  preferred the non-supersymmetric vacuum,''
  JHEP {\bf 0701}, 089 (2007)
  [arXiv:hep-th/0610334].
  %%CITATION = JHEPA,0701,089;%%
}

%\CraigKX
\lref\CraigKX{
  N.~J.~Craig, P.~J.~Fox and J.~G.~Wacker,
  ``Reheating metastable O'Raifeartaigh models,''
  Phys.\ Rev.\  D {\bf 75}, 085006 (2007)
  [arXiv:hep-th/0611006].
  %%CITATION = PHRVA,D75,085006;%%
}

%\FischlerXH
\lref\FischlerXH{
  W.~Fischler, V.~Kaplunovsky, C.~Krishnan, L.~Mannelli and M.~A.~C.~Torres,
  ``Meta-Stable Supersymmetry Breaking in a Cooling Universe,''
  JHEP {\bf 0703}, 107 (2007)
  [arXiv:hep-th/0611018].
  %%CITATION = JHEPA,0703,107;%%
}

%\KutasovKB
\lref\KutasovKB{
  D.~Kutasov, O.~Lunin, J.~McOrist and A.~B.~Royston,
  ``Dynamical Vacuum Selection in String Theory,''
  Nucl.\ Phys.\  B {\bf 833}, 64 (2010)
  [arXiv:0909.3319 [hep-th]].
  %%CITATION = NUPHA,B833,64;%%
}

%\BinetruyXJ
\lref\BinetruyXJ{
  P.~Binetruy and G.~R.~Dvali,
  ``D-term inflation,''
  Phys.\ Lett.\  B {\bf 388}, 241 (1996)
  [arXiv:hep-ph/9606342].
  %%CITATION = PHLTA,B388,241;%%
}

%\HalyoPP
\lref\HalyoPP{
  E.~Halyo,
  ``Hybrid inflation from supergravity D-terms,''
  Phys.\ Lett.\  B {\bf 387}, 43 (1996)
  [arXiv:hep-ph/9606423].
  %%CITATION = PHLTA,B387,43;%%
}

%\KachruGS
\lref\KachruGS{
  S.~Kachru, J.~Pearson and H.~L.~Verlinde,
  ``Brane/Flux Annihilation and the String Dual of a Non-Supersymmetric Field
  Theory,''
  JHEP {\bf 0206}, 021 (2002)
  [arXiv:hep-th/0112197].
  %%CITATION = JHEPA,0206,021;%%
}

%\WittenMT
\lref\WittenMT{
E.~Witten,
  ``Solutions of four-dimensional field theories via M theory,''
  Nucl.\ Phys.\ B {\bf 500}, 3 (1997) 
  [arXiv:hep-th/9703166].
  %%CITATION=NUPHA,B500,3;%%
}

%\BenaXK
\lref\BenaXK{
  I.~Bena, M.~Grana and N.~Halmagyi,
  ``On the Existence of Meta-stable Vacua in Klebanov-Strassler,''
JHEP {\bf 1009}, 087 (2010).
[arXiv:0912.3519 [hep-th]].
%%CITATION = arXiv:0912.3519%%
}

%\DymarskyPM
\lref\DymarskyPM{
  A.~Dymarsky,
  ``On gravity dual of a metastable vacuum in Klebanov-Strassler theory,''
JHEP {\bf 1105}, 053 (2011).
[arXiv:1102.1734 [hep-th]].
%%CITATION = arXiv:1102.1734%%
}

%\BlabackNF
\lref\BlabackNF{
  J.~Blaback, U.~H.~Danielsson and T.~Van Riet,
  ``Resolving anti-brane singularities through time-dependence,''
[arXiv:1202.1132 [hep-th]].
%%CITATION = arXiv:1202.1132%%
}

%\MassaiJN
\lref\MassaiJN{
  S.~Massai,
  ``A Comment on anti-brane singularities in warped throats,''
[arXiv:1202.3789 [hep-th]].
%%CITATION = arXiv:1202.3789%%
}

%\PolchinskiMX
\lref\PolchinskiMX{
  J.~Polchinski,
  ``N = 2 gauge-gravity duals,''
  Int.\ J.\ Mod.\ Phys.\  A {\bf 16}, 707 (2001)
  [arXiv:hep-th/0011193].
  %%CITATION = IMPAE,A16,707;%%
}

%\BeniniIR
\lref\BeniniIR{
  F.~Benini, M.~Bertolini, C.~Closset and S.~Cremonesi,
  ``The N=2 cascade revisited and the enhancon bearings,''
  Phys.\ Rev.\  D {\bf 79}, 066012 (2009)
  [arXiv:0811.2207 [hep-th]].
  %%CITATION = PHRVA,D79,066012;%%
}

%\UrangaVF
\lref\UrangaVF{
  A.~M.~Uranga,
  ``Brane configurations for branes at conifolds,''
JHEP {\bf 9901}, 022 (1999).
[hep-th/9811004].
%%CITATION = hep-th/9811004%%
}

%\DasguptaSU
\lref\DasguptaSU{
  K.~Dasgupta and S.~Mukhi,
  ``Brane constructions, conifolds and M theory,''
Nucl.\ Phys.\ B {\bf 551}, 204 (1999).
[hep-th/9811139].
%%CITATION = hep-th/9811139%%
}

%\DasguptaSW
\lref\DasguptaSW{
K.~Dasgupta, J.~Seo, A.~Wissanji
  ``F-theory, Seiberg-Witten Curves and $N=2$ Dualities,''
  JHEP {\bf 1202}, 146 (2012) 
  [arXiv:1107.3566[hep-th]].
  %%CITATION=JHEPA,1202,146;%%
}

%\ArgyresEH
\lref\ArgyresEH{
  P.~C.~Argyres, M.~R.~Plesser and N.~Seiberg,
  ``The Moduli space of vacua of N=2 SUSY QCD and duality in N=1 SUSY QCD,''
Nucl.\ Phys.\ B {\bf 471}, 159 (1996).
[hep-th/9603042].
%%CITATION = hep-th/9603042%%
}

%%%%%%%%%%%%%%%%%%%%%%%%%%%%%%%%%%%%%%%%%%%%%%%%%%%
\Title{} {\centerline{IIA perspective on Cascading Gauge Theory}}

\bigskip
\centerline{\it David Kutasov $^1$ and Alisha Wissanji $^2$}
\bigskip

\centerline{${}^{1}$EFI and Department of Physics, University of
Chicago}\centerline{5640 S. Ellis Av. Chicago, IL 60637}\centerline{dkutasov@uchicago.edu}
\bigskip
\centerline{${}^{2}$Department of Physics, McGill University} 
\centerline{3600 University Street, Montreal QC, Canada H3A 2T8}
\centerline{wissanji@hep.physics.mcgill.ca}
\smallskip

\vglue .3cm

\bigskip

\bigskip
\noindent

We study the $N=1$ supersymmetric cascading gauge theory found in type IIB string theory on $p$ regular and $M$ fractional $D3$-branes at the tip of the conifold, using the T-dual type IIA description. We reproduce the supersymmetric vacuum structure of this theory, and show that the IIA analog of the non-supersymmetric state found by Kachru, Pearson and Verlinde in the IIB description is metastable in string theory, but the barrier for tunneling to the supersymmetric vacuum goes to infinity in the field theory limit.  We also comment on the $N=2$ supersymmetric gauge theory corresponding to regular and fractional $D3$-branes on a near-singular $K3$, and clarify the origin of the cascade in this theory.

\bigskip

\Date{}

%%%%%%%%%%%%%%%%%%%%%%%%%%%%%%%%%%%%%%%%%%%%%%%%%%%%%%%%%%%%%%%%
%%%%%%%%%%%%%%%%%%%%%%%%%%%%%%%%%%%%%%%%%%%%%%%%%%%%%%%%%%%%%%%%

\newsec{Introduction}

The system of $p$ $D3$-branes and $M$ $D5$-branes at the tip of the conifold 
in type IIB string theory \KlebanovHB\  exhibits many non-trivial phenomena such as 
confinement, dynamical symmetry breaking and a rich landscape of ground states. 
It is also an important example of gauge/gravity duality, which plays a role in studies 
of string phenomenology and early universe cosmology. 

The low energy effective field theory of this system is an $N=1$ supersymmetric 
four dimensional gauge theory with gauge group $SU(M+p)\times SU(p)$ and 
matter in the bifundamental representation \KlebanovHB. The rich vacuum structure 
of this gauge theory was described in \DymarskyXT. It can be interpreted in terms of 
a ``duality cascade'' -- a sequence of gauge theories with varying ranks which provide 
a description of the different vacua. Some of these vacua have a regular type IIB  
supergravity description, which has also been extensively investigated. 

The T-dual of the type IIB  construction of \KlebanovHB\ is given by a system 
of $NS5$-branes and $D4$-branes in type IIA string theory. This system was 
mentioned in \KlebanovHB\ and further studied in \AharonyMI. One of our goals 
below will be to build on the results of \AharonyMI\ and reproduce the results of 
\refs{\KlebanovHB,\DymarskyXT} using the IIA description. We will also discuss 
some non-supersymmetric aspects of the dynamics. 

We will see that the IIA description provides a nice picture of the supersymmetric 
and non-supersymmetric vacua. As is standard in studying
brane dynamics in string theory, the three descriptions (gauge theory, IIA and IIB) are valid in
different regions in the parameter space of the brane system. This should not matter for the
supersymmetric vacuum structure, and indeed we will reproduce the results of 
\refs{\KlebanovHB,\DymarskyXT} in the IIA language. Many aspects of the 
non-supersymmetric vacuum structure are also expected to agree, and we will find 
that to be the case.  

Cascading behavior was found in a wide variety of theories, some of which do not exhibit Seiberg 
duality. We will briefly discuss an example of this phenomenon, an $N=2$ supersymmetric quiver theory
closely related to $N=2$ SQCD, and use the IIA description to identify the origin of the cascade in this theory. 

The plan of the paper is as follows. In section 2 we introduce the classical $N=1$ supersymmetric 
gauge theory and IIA brane system that reduces to it at low energies. We review the structure of the 
classical moduli spaces in both languages, and show that they agree. We also discuss some non-supersymmetric 
vacua that appear for non-zero Fayet-Iliopoulos (FI) coupling. 

In section 3 we discuss the quantum theory. We show that the quantum moduli space of the brane system
is the same as that of the gauge theory, and in particular exhibits the cascading behavior found in
\refs{\KlebanovHB,\DymarskyXT}. The brane picture gives a simple description of the cascade and helps 
understand which vacua of theories with different values of $p$ agree, and which do not. In this
picture, the cascade is associated with the fact that for a given value of the UV cutoff the fivebrane in general 
winds around a circle. As one reduces the UV cutoff, the winding number decreases. This 
corresponds in the field theory language to decreasing $p$ by a multiple of $M$. Vacua in which the fivebrane
does not wind (or does not wind enough times) around the circle are not in general the same in theories with 
different values of $p$. 

In section 4 we discuss non-supersymmetric vacua of the brane system. We show that for the non-supersymmetric vacua that appear for non-zero FI coupling the quantum brane picture incorporates chiral symmetry breaking, which is expected to occur in the corresponding low energy gauge theory. We also discuss the IIA analog of the metastable states discussed in the IIB language in \KachruGS. 
We find that these states are present in the brane system, but the barrier that separates them from the supersymmetric states goes to infinity in the field theory limit. Thus, they become stable in that limit. 

In section 5 we discuss the $N=2$ supersymmetric analog of the cascading gauge theory, and in particular address the question how a theory that does not have Seiberg duality can have a duality cascade. We show that the situation is similar to that in the $N=1$ case -- the gauge theory has a rich set of vacua, some of which exhibit cascading behavior.  Even in these vacua, different theories along the cascade differ by abelian factors in the gauge group. 

Section 6 contains a brief discussion of our results; an appendix summarizes some aspects of the IIA description of quantum $N=2$ SQCD, which are useful for the discussion in section 5.

\newsec{Classical theory}

We start with a brief description of the classical gauge theory and the corresponding type 
IIA brane system. We refer the reader to \refs{\DymarskyXT,\GiveonSR} for a more detailed 
discussion of the two topics. We will draw heavily on the results described in these papers.

As mentioned in the introduction, we will be studying an $N=1$ supersymmetric gauge theory with gauge group 
\eqn\ggroup{G=SU(N_1)\times SU(N_2)~,}
with 
\eqn\defnonetwo{N_1=M+p;\qquad N_2=p~.}
The matter consists of chiral superfields $A_{\alpha i}^a$ and $B_{\dot\alpha a}^i$,
where $i=1,\cdots, N_1$, $a=1,\cdots, N_2$, are gauge indices, and $\alpha,\dot\alpha=1,2$
are global symmetry labels. As implied by the notation, the matter fields transform under
$G$ as follows:
\eqn\transfab{\eqalign{A_\alpha &\qquad (N_1,\bar N_2)~,\cr
B_{\dot\alpha} &\qquad (\bar N_1, N_2)~.
}}
There is also a tree level superpotential,
\eqn\www{W_0={h\over2}\epsilon^{\alpha\beta}\epsilon^{\dot\alpha\dot\beta}
A_{\alpha i}^aB_{\dot\alpha b}^iA_{\beta j}^bB_{\dot\beta a}^j=
h\left(A_{1 i}^aB_{1 b}^iA_{2 j}^bB_{2 a}^j-A_{1 i}^aB_{2 b}^iA_{2 j}^bB_{1 a}^j\right).
}
To compare to standard discussions of $N=1$ SQCD, it is useful to note 
that the $SU(N_1)$ factor in the gauge group \ggroup\ ``sees'' $2N_2$ flavors,
and similarly for $SU(N_2)$. Since $N_1>N_2$ \defnonetwo, in the quantum 
theory the quartic superpotential \www\ is a relevant perturbation of the IR fixed 
point of the $SU(N_1)$ gauge theory obtained  by turning off the $SU(N_2)$ 
gauge coupling, and an irrelevant perturbation of the corresponding $SU(N_2)$ fixed point. 
 
The global symmetry of the model includes $SU(2)\times SU(2)$, with the
two factors acting on the indices $\alpha$ and $\dot\alpha$, and a $U(1)$ 
symmetry which assigns charges $+1$ to $A$ and $-1$ to $B$; this symmetry
is usually referred to as baryon number. We will denote it by $U(1)_b$
and will mostly consider the  theory in which it is gauged, since this is the case 
in the IIA brane system we will study. It is also useful to consider this case in the  
IIB theory, since it is relevant to the embedding of the conifold geometry in a compact 
Calabi-Yau manifold. 

\bigskip

\ifig\loc{The IIA brane configuration that realizes the cascading gauge theory.  
The $NS5$-branes are depicted in green, and are connected by $M+p$ 
$D4$-branes on one side of the $x^6$ circle (whose radius is $R_6$) and
by $p$ $D4$-branes on the other.}
{\epsfxsize3.3in\epsfbox{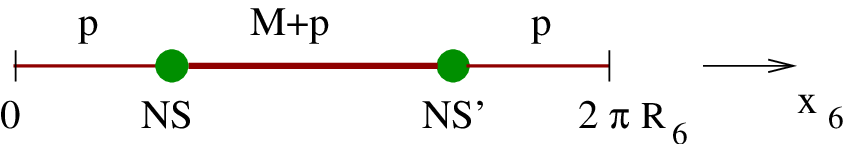}}

\bigskip

The IIA brane system that gives rise to the above gauge theory at low energies is
depicted in figure 1. The system contains two kinds of branes: $NS5$-branes 
localized on a circle $x^6\sim x^6+2\pi R_6$, represented by green circles in the figure, and stacks of 
$D4$-branes connecting them, represented by red lines. The orientations of the 
different branes in the $9+1$ dimensional spacetime are as follows:
\eqn\braneor{\eqalign{
NS: \qquad & 012345~,\cr
NS': \qquad & 012389~,\cr
D4: \qquad & 01236~.\cr}}
This configuration preserves $N=1$ SUSY in the $3+1$ dimensions common to all
the branes, $(0123)$. 

The low energy effective theory of this brane system contains $U(M+p)\times U(p)$ 
$N=1$ SYM, associated with the massless excitations living on the $M+p$ and $p$
$D4$-branes, respectively, and  chiral superfields $A, B$  \transfab, which come from
strings connecting the two stacks of $D4$-branes. An overall $U(1)$ in 
$U(M+p)\times U(p)$ is decoupled, and can be ignored, but the relative $U(1)$ is
precisely the $U(1)_b$ discussed above. Hence, the brane construction
gives the theory in which this symmetry is gauged, as mentioned above. 

The classical $U(N_i)$ gauge couplings are determined by the lengths of the corresponding
branes, $L_i$,
\eqn\gili{{1\over g_i^2}={L_i\over g_sl_s}~.}
As is clear from figure 1,
\eqn\sumlength{L_1+L_2=2\pi R_6,\qquad i.e. \qquad 
{1\over g_1^2}+{1\over g_2^2}={2\pi R_6\over g_sl_s}.}
Our purpose in the remainder of this section is to compare the classical moduli space
of supersymmetric vacua of the gauge theory, studied in \DymarskyXT,  to that of the 
brane system. We will also discuss some non-supersymmetric vacua of the theory.
In the next section we will describe the quantum moduli space.

The D-term equations of this gauge theory can be written as the following matrix equations
for the $(p+M)\times p$ matrices $A_\alpha$, $B_{\dot\alpha}^\dagger$:
\eqn\dterms{\eqalign{
\sum_\alpha A_\alpha A^\dagger_\alpha-\sum_{\dot\alpha}B^\dagger_{\dot\alpha} B_{\dot\alpha}=&
{\CU\over p} I_p~,\cr
\sum_\alpha A^\dagger_\alpha A_\alpha-\sum_{\dot\alpha}B_{\dot\alpha} B^\dagger_{\dot\alpha}=&
{\CU\over M+p} I_{M+p}~,\cr}}
with $I_n$ an $n\times n$ identity matrix, and
\eqn\defuu{\CU={\rm Tr}\left(\sum_\alpha A_\alpha A^\dagger_\alpha-
\sum_{\dot\alpha}B^\dagger_{\dot\alpha} B_{\dot\alpha}\right)~.}
In the theory with gauged $U(1)_b$, one must set $\CU=0$; turning on a 
Fayet-Iliopoulos (FI) term $\xi$ for this $U(1)$, modifies this to 
\eqn\uxi{\CU=\xi~.}
Classical supersymmetric vacua correspond to solutions of the D-term equations \dterms\ -- \uxi\ as well
as the F-term conditions for the superpotential \www. For general $M$, $p$, and setting $\xi=0$ for now,
the solutions of these equations can be written, up to gauge transformations, in the diagonal form
\eqn\formab{\eqalign{
A_\alpha=&\left(\matrix{A_{\alpha1}^1& & & & & & & &\cr
 &A_{\alpha2}^2& & & & & & &\cr
 & &A_{\alpha3}^3& & & & &\cr
 & & &.& & & & &\cr
 & & & &. & & & &\cr
 & & & & &A_{\alpha p}^p& & \cr}\right)~,\cr
B^{t}_{\dot\alpha}=&\left(\matrix{B_{\dot\alpha1}^1& & & & & & & &\cr
 &B_{\dot\alpha2}^2& & & & & & &\cr
 & &B_{\dot\alpha3}^3& & & & &\cr
 & & &.& & & & &\cr
 & & & &.& & & &\cr
 & & & & &B_{\dot\alpha p}^p&  & \cr}\right)~.}}
The eigenvalues  $A_{\alpha a}^a$ and  $B_{\dot\alpha a}^a$, $a=1,\cdots, p$ satisfy the constraints 
\eqn\dconst{\sum_\alpha|A_{\alpha a}^a|^2-\sum_{\dot\alpha}|B_{\dot\alpha a}^a|^2=0~.}
For given $a$, the eigenvalues are four complex fields, which satisfy one real constraint  
\dconst. Another real field (for each $a$) is removed by the (Higgsed) gauge symmetry. 
Thus, the moduli space is $3p$ (complex) dimensional. It can be described by the $4p$ 
complex coordinates 
\eqn\zzaa{z^a_{\alpha\dot\alpha}=A_{\alpha a}^aB_{\dot\alpha a}^a~,}
which satisfy the (complex) constraints 
\eqn\singcon{\det_{\alpha\dot\alpha} z^a_{\alpha\dot\alpha}=0~.}
Together with the symmetry of permutation of the $p$ eigenvalues, we conclude that 
the classical moduli space is a symmetric product of $p$ copies of the singular conifold 
\singcon,
\eqn\clasmod{\MM_0={\rm Sym}_p(\CC_0)~.}
At a generic point in the moduli space, the low energy theory consists of an $SU(M)$ $N=1$ 
SYM theory\foot{The unbroken $SU(M)$ is a  subgroup of the $SU(M+p)$ factor in  \ggroup.} 
and  $p$ copies  of $N=4$ SYM with gauge group $U(1)$.

In terms of the brane system of figure 1, the moduli space described above is obtained by 
noting that the configuration contains $p$ $D4$-branes that wrap the circle, and can thus 
freely move in the $\IR^5$ labeled by $(45789)$; another (compact) dimension 
of moduli space is obtained from a component of the gauge field on the fourbanes, $A^6$. 
A generic point in the moduli space is described in figure 2.  The $p$ mobile branes support 
$U(1)^p$ $N=4$ SYM, while the $M$ localized branes give rise to pure $N=1$ SYM with 
gauge group $SU(M)$ (and a decoupled $U(1)$ mentioned above), in agreement with the 
gauge theory analysis.

\bigskip

\ifig\loc{A generic point in the classical moduli space. $p$ $D4$-branes wrap the $x^6$ circle 
and can move in the transverse space; $M$ are stretched between the fivebranes and give 
rise to $SU(M)$ $N=1$ SYM.}
{\epsfxsize3.4in\epsfbox{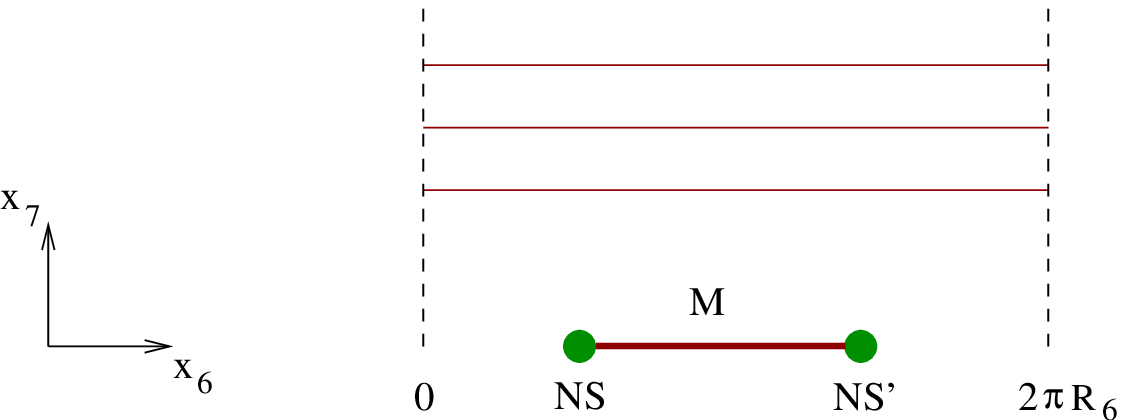}}

\bigskip

The form of the moduli space 
\clasmod\ is very natural from the brane perspective: in the classical gauge theory limit, the 
separation between the fivebranes goes to zero \GiveonSR, and the $M$ $D4$-branes in 
figure 2 can be ignored. The fivebranes are described by the equation  $vw=0$, where 
\eqn\vvww{v=x^4+i x^5~;\qquad w=x^8+i x^9~.}
This is known to be a dual description of the conifold (obtained by T-duality in $x^6$; see \eg\ 
\refs{\UrangaVF\DasguptaSU-\McOristIN}). Under this T-duality, the mobile $D4$-branes turn into 
$D3$-branes living on the conifold, in agreement with \clasmod. 
 
We next turn to the case where the FI parameter $\xi$ \uxi\ is non-vanishing. In general, supersymmetry 
is then broken, with vacuum energy $V\sim g^2\xi^2$ \DymarskyXT. The case 
\eqn\pkm{p=kM,}
with integer $k$ is special. In that case the gauge theory has an isolated supersymmetric vacuum, 
in which for $\xi>0$ one has $B_{\dot\alpha}=0$, 
\eqn\susyaone{A_{\alpha=1}=C\left(\matrix{\sqrt{k}&0&0&.&0&0\cr
0&\sqrt{k-1}&0&.&0&0\cr
0&0&\sqrt{k-2}&.&0&0\cr
.&.&.&.&.&.\cr
0&0&0&.&1&0\cr}\right),}
and 
\eqn\susyatwo{A_{\alpha=2}=C\left(\matrix{0&1&0&.&0&0\cr
0&0&\sqrt{2}&.&0&0\cr
0&0&0&\sqrt{3}&.&0\cr
.&.&.&.&.&.\cr
0&0&0&.&0&\sqrt{k}\cr}\right)~.}
Each entry in the matrices \susyaone, \susyatwo\ is proportional to an $M\times M$ 
unit matrix, and the constant $C$ satisfies \defuu, \uxi, $\xi=k(k+1)M|C|^2$. For 
$\xi<0$, one finds a similar vacuum with $A\leftrightarrow B$. The low energy theory 
in the baryonic vacuum is described by an unbroken $SU(M)$ $N=1$ SYM, but unlike
the mesonic branch, this $SU(M)$ is embedded non-trivially in both factors of the
gauge group $G$ \ggroup. 
 
In the theory with gauge group  $SU((k+1)M)\times SU(kM)$ 
(\ie\ with ungauged baryon number), \susyaone, \susyatwo\ give rise to a one complex
dimensional moduli space of vacua labeled by $C$, which is usually refered to as the
baryonic branch. Our interest is in the theory where $U(1)_b$ is gauged, 
in which (classically) it only appears for non-zero $\xi$ and is isolated. 

In the brane system, the FI coupling $\xi$ corresponds to the relative displacement 
between the fivebranes in $x^7$ \GiveonSR. It is clear that generically this breaks 
supersymmetry, leading to configurations such as that of figure 3, in which different 
$D4$-branes  are not mutually BPS. 
 
\bigskip

\ifig\loc{Turning on an FI term in general leads to branes at an angle and breaks SUSY.}
{\epsfxsize4.0in\epsfbox{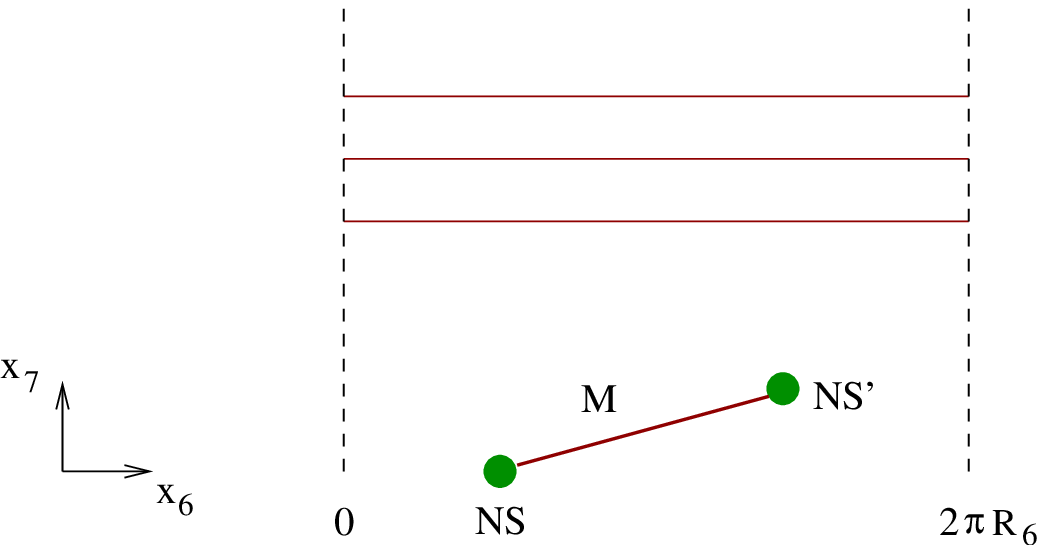}}

\bigskip

The baryonic vacuum \susyaone, \susyatwo\ is described in terms of the branes by the 
configuration of figure 4. The red line corresponds to a stack of $M$ $D4$-branes, which
connects the $NS$ and $NS'$-branes, in the process winding $k$ times around the circle. 
It is easy to check that all the branes in figure 4 are mutually BPS and the configuration 
is supersymmetric.  Note that the vacuum of figure 4 is isolated, as expected from the 
gauge theory analysis. Turning off the FI term, \ie\ taking the $NS'$-brane in figure 4 to the
 $x^6$ axis, leads to the configuration of  figure 1, with $p=kM$. Thus, the baryonic vacuum 
 coincides in this case with the origin of the mesonic branch, as in gauge theory.

\bigskip

\ifig\loc{The baryonic vacuum for $\xi\not=0$ corresponds to a stack of $M$ $D4$-branes
connecting the fivebranes and winding $k$ times around the circle ($k=2$ in the figure). }
{\epsfxsize4.0in\epsfbox{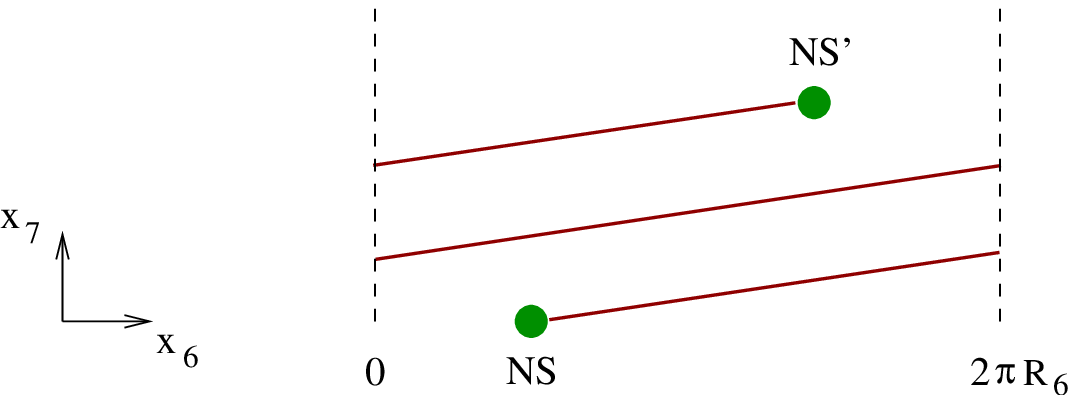}}

\bigskip

It is also clear from the figure that the low energy theory in this vacuum is pure $N=1$ SYM 
with gauge group $SU(M)$, and that this $SU(M)$ is non-trivially embedded in the full gauge 
symmetry  $SU((k+1)M)\times SU(kM)$. A quick way to find this embedding is to calculate
the gauge coupling of the unbroken $SU(M)$, $g$. Taking $\xi\to 0$, the 
coupling is related to the length of the branes, as in \gili:
\eqn\couplbar{{1\over g^2}={L_1+2\pi kR_6\over g_s l_s}={k+1\over g_1^2}+{k\over g_2^2}~.} 
Thus, the unbroken gauge group is the diagonal $SU(M)$ subgroup of an $SU(M)^{k+1}$ subgroup
in $SU((k+1)M)$ and an $SU(M)^k$ subgroup of $SU(kM)$,  in agreement with the gauge theory. 

Finally, figure 4 makes it clear that by moving the $NS'$-brane in $x^6$, we can 
change the winding number of the spiraling $D4$-branes, and thus $k$, by one 
or more units, without changing the low energy theory.\foot{Of course, if we keep 
the parameters $L$, $R_6$ fixed in the process, the (classical) gauge coupling of 
the $SU(M)$ gauge theory changes, but we can adjust these parameters so that 
it does not.}  This is a classical precursor of  Seiberg duality \SeibergPQ, which is known to play 
an important role in the quantum dynamics of the cascading gauge theory. The way
 it appears here is reminiscent of the discussion of \ElitzurFH. We will discuss its
 quantum analog in the next section.

In addition to the supersymmetric vacuum of figure 4 (or eqs \susyaone, \susyatwo\ in gauge 
theory) the brane system has a series of non-supersymmetric vacua labeled  by the winding 
number of the $M$ $D4$-branes stretched between the fivebranes, $l=0,1,2,\cdots, k$. The 
vacuum with winding number $l$ contains $M$ $D4$-branes connecting the $NS$ and 
$NS'$-branes while winding $l$ times around the circle, and $p-lM$ mobile $D4$-branes.  
The vacuum with $l=0$ is the one described in figure 3, while that with $l=k$ corresponds to figure 4 
(and is supersymmetric for $p=kM$). 

Since the vacua with $l<k$ are not supersymmetric, it is natural to ask what is the potential 
on the $3(p-lM)$ complex dimensional pseudomoduli space. Far along the moduli space (\ie\
for large $A$, $B$ in \dconst, or $z$ in \zzaa) and at weak IIA string coupling, it is clear 
that the leading effect is due to closed string exchange between the mobile fourbranes and those
stretching between the fivebranes. Since these branes are not parallel, the gravitational attraction 
does not precisely cancel the RR repulsion, and there is a net attractive force pulling the mobile
branes towards the localized ones. 

The resulting dynamics facilitates a change in $l$, as demonstrated in figure 5, in which a stack of 
$M$ mobile $D4$-branes is pulled towards the fivebranes (a); when it intersects them (b), the brane 
configuration has an instability towards reconnection (c), due to the presence of an 
open string tachyon living at the intersection. Condensation of this tachyon leads to the 
configuration (d), in which $l$ has increased by one unit. The endpoint of this process is the 
supersymmetric vacuum, with $l=k$ (figure 4).\foot{One might think that there are other non-supersymmetric
but locally stable ground states in which the winding numbers of the different fourbranes connecting the fivebranes 
are different and the $SU(M)$ gauge symmetry is broken, but this is not the case.}

\bigskip

\ifig\loc{Open string tachyon condensation connects vacua with different values of $l$. }
{\epsfxsize4.5in\epsfbox{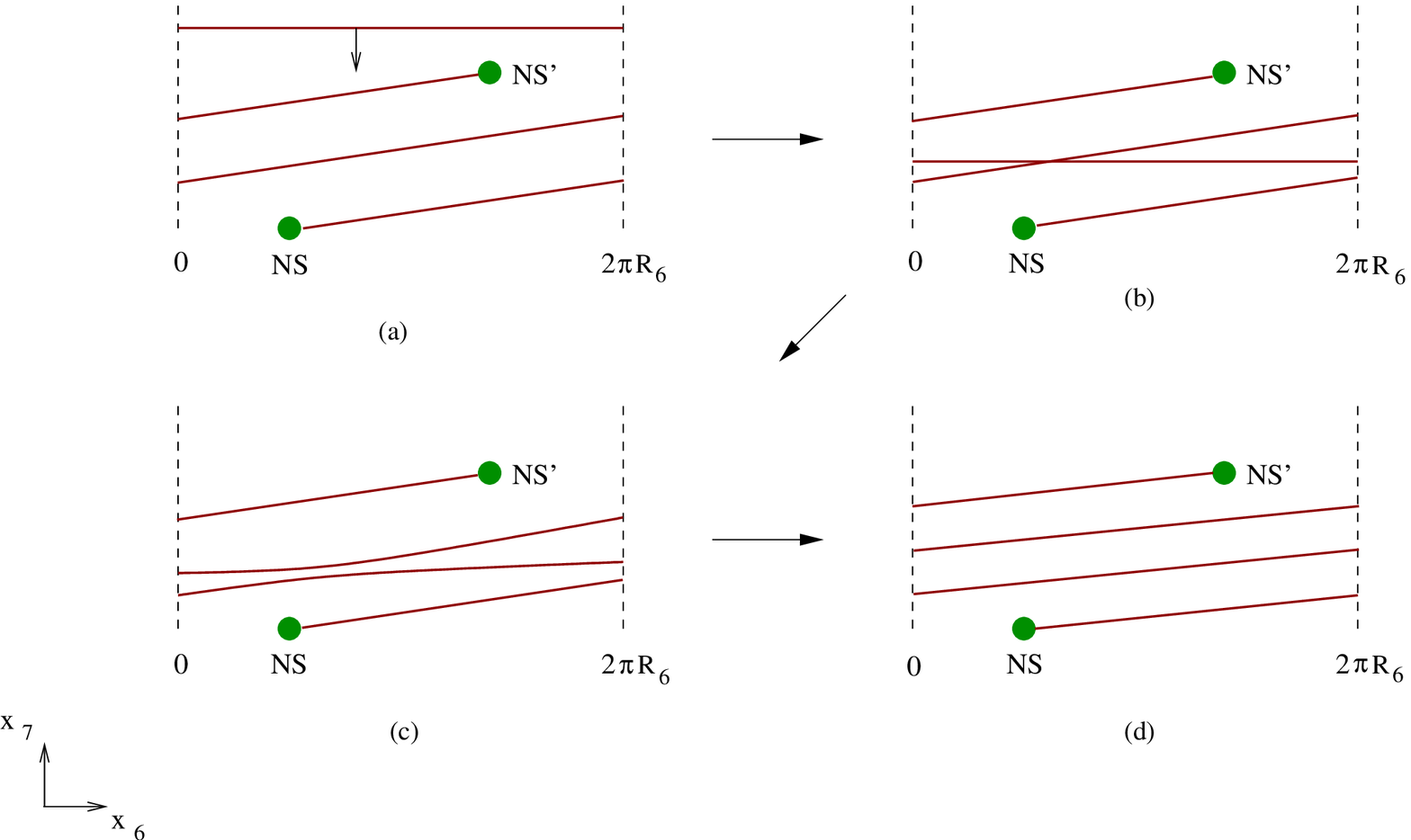}}

\bigskip

The above brane discussion has a gauge theory counterpart. The F and D term potential at 
non-zero $\xi$ has a series of non-supersymmetric vacua in which the matrices $A$ and $B$
split into a block of size $(l+1)M\times lM$ in which they look like \susyaone, \susyatwo\ with 
$k\to l$, and a block of size $p-lM$, which looks like  \formab. The eigenvalues \dconst\ label 
the pseudomoduli space as in the discussion around \zzaa. The potential for the pseudomoduli
is classically flat,  however near the origin of pseudomoduli space there is a tachyonic instability 
in a different direction in field space, which takes the system towards the supersymmetric vacuum 
\susyaone, \susyatwo. 

A natural question is what is the field theory analog of the classical gravitational attraction that in 
the brane description gives a potential on pseudomoduli space and leads to the isolated baryonic 
vacuum of figure 4. In other closely related brane systems, such as those that appear in the discussion 
of the ISS model, this potential is the Coleman-Weinberg (CW) potential computed in \IntriligatorDD. 
It is natural to expect the same to happen here; we will leave a detailed analysis to future work. 

The gravitational brane attraction can be described from the field theory point of view in terms of 
a non-canonical Kahler potential for the light fields. As disussed in 
\refs{\GiveonFK\GiveonEW-\GiveonUR}, this effect is not identical to the CW potential. The two
are dominant in different regions in the parameter space of the brane system,  but tend 
to lead to similar dynamics. 

So far we focused on the case $p=kM$ \pkm, but it is easy to generalize to
\eqn\ptilde{p=kM+\tilde p;\qquad 1\le \tilde p\le M-1~.}
Most of the discussion of this case is the same as before. After all but $\tilde p$ of the mobile 
$D4$-branes have combined with the localized fourbranes via the process of figure 5, we are 
left with $\tilde p<M$ branes in the bulk. These branes are also attracted to the spiraling fourbrane
and undergo a process similar to that of figure 5, except now it affects only $\tilde p$
of the $M$ spiraling fourbranes. This leads to a state in which we have $M-\tilde p$ 
fourbranes which stretch between the fivebranes while winding $k$ times around the 
circle, and $\tilde p$ fourbranes which wind $k+1$ times (see figure 6). Clearly, this state is not
supersymmetric (for generic $\tilde p$). We see that in this case, turning on an FI term causes
the moduli space to collapse to an isolated non-supersymmetric vacuum.

\bigskip

\ifig\loc{The ground state of the brane system with non-zero $\xi$, viewed in the covering space of
the $x^6$ circle. $M-\tilde p$ fourbranes have winding $k$, while $\tilde p$ have winding $k+1$.
The specific case exhibited is  $k=\tilde p=1$, $M=2$, $p=3$. }
{\epsfxsize4.5in\epsfbox{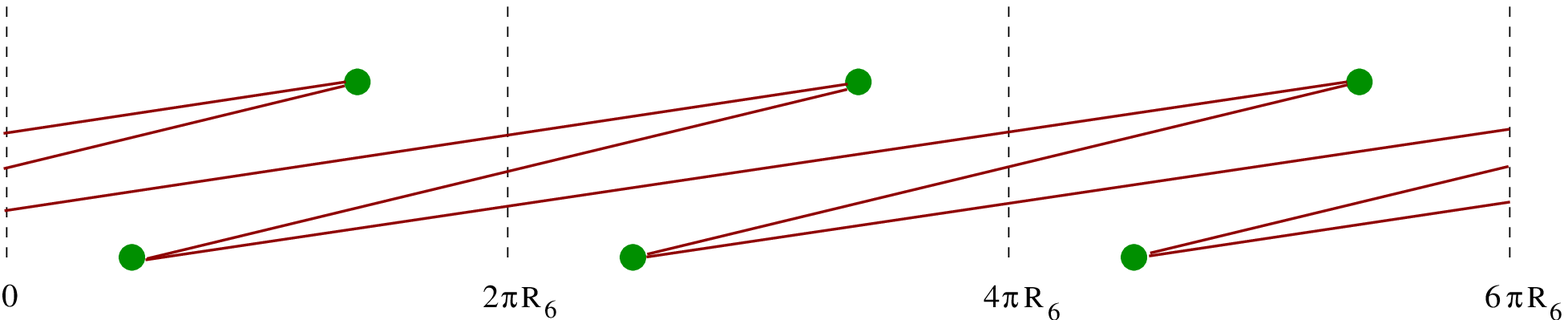}}

\bigskip

The low energy dynamics of the theory with non-zero $\tilde p$ can be read off from figure 6. 
The unbroken gauge group is $SU(M-\tilde p)\times SU(\tilde p)\times U(1)_b$. 
The embedding of this group in \ggroup\ can be determined in a similar way to the discussion 
around \couplbar. The gauge group can be written as 
\eqn\gaugtildep{SU(M+p)\times SU(p)=SU\left((k+1)(M-\tilde p)+(k+2)\tilde p\right)\times SU(k(M-\tilde p)+
(k+1)\tilde p)~.}
The first factor contains a $\left[SU(M-\tilde p)\right]^{k+1}\times \left[SU(\tilde p)\right]^{k+2}$ subgroup;
the second contains $\left[SU(M-\tilde p)\right]^k\times \left[SU(\tilde p)\right]^{k+1}$. The gauge group 
corresponding to figure 6 involves the diagonal $SU(M-\tilde p)\times SU(\tilde p)$ of all these factors.

There are two kinds of matter fields. One comes from strings both of whose ends lie on the same stack 
of fourbranes. In addition to the gauge fields, these give fermions in the adjoint representation of the gauge
group, which in the absence of the second stack of fourbranes would be the gauginos of an $N=1$ 
supersymmetric model. The second comes from open strings stretched between the two stacks, and is 
localized at their intersections. As shown in \gkunpub,  two $D4$-branes ending on an $NS5$-brane at a 
generic angle give rise to a massless Dirac fermion.\foot{The lightest bosonic fields have a mass that 
depends on the angle between the fourbranes and is non-zero unless these branes are parallel.}
Thus, the vacuum of the theory of figure 6 contains fermions in the 
bifundamental of  $SU(M-\tilde p)\times SU(\tilde p)$ charged under $U(1)_b$, \ie\ the
light matter is similar to that of the original cascading gauge theory, without the scalars. 

So far we discussed the non-supersymmetric vacuum of the theory with generic $\tilde p$ from the point
of view of the IIA brane construction, but it is easy to repeat the discussion in the gauge theory language. 
For $\xi>0$, the vacuum field configuration is obtained by splitting each $M\times M$ block on the diagonal 
in \susyaone, \susyatwo\ into 
blocks of size $M-\tilde p$ and $\tilde p$. Looking back at \gaugtildep\ we see that in the blocks of size 
$M-\tilde p$ we should use the ansatz \susyaone, \susyatwo, with $C=C_{M-\tilde p}$ . In the blocks of
size $\tilde p$ we should use a similar ansatz, with $k\to k+1$ and $C=C_{\tilde p}$. The D-term potential
takes the form (up to an overall constant)
\eqn\dtermpot{\eqalign{
V_D\simeq & k(M-\tilde p)\left(|C_{M-\tilde p}|^2(k+1)-{\xi\over p}\right)^2+
(k+1)\tilde p\left(|C_{\tilde p}|^2(k+2)-{\xi\over p}\right)^2\cr
+ & (k+1)(M-\tilde p)\left(|C_{M-\tilde p}|^2k-{\xi\over p+M}\right)^2+
(k+2)\tilde p\left(|C_{\tilde p}|^2(k+1)-{\xi\over p+M}\right)^2~.\cr
}}
Minimizing w.r.t. $C_{\tilde p}$ and $C_{M-\tilde p}$ we find 
\eqn\cmp{\eqalign{|C_{M-\tilde p}|^2=&{\xi\over 2k+1}\left({1\over p}+{1\over M+p}\right),\cr
|C_{\tilde p}|^2=&{\xi\over 2k+3}\left({1\over p}+{1\over M+p}\right).\cr
}}
For $\tilde p=0,M$ this reduces to the supersymmetric result of \DymarskyXT. 

To summarize, we found that the classical brane configuration has the same vacuum structure as 
the classical gauge theory. As usual \GiveonSR, the IIA description provides a simple geometric 
picture of the vacuum structure and low energy dynamics in a certain region of the parameter 
space of brane configurations. In the next section we move on to the quantum theory and compare 
the structure one finds in the gauge theory and brane pictures.

\newsec{Quantum theory}

In studying quantum effects we start from small values of $p$, and then proceed to larger ones. 

\subsec{$p=0$}

The field theory  described in section 2 is in this case $N=1$ pure SYM with gauge group $SU(M)$.
This theory generates dynamically a mass gap $\Lambda_1$,  and has $M$ isolated vacua in which 
the  superpotential takes the values 
\eqn\dynsup{W=M\Lambda_1^3e^{2\pi i r\over M};\qquad r=1,\cdots, M~.}
The index $r$ labels $M$ vacua related by a $Z_{2M}$ R-symmetry (the anomaly free part of a $U(1)_R$
symmetry), which is dynamically broken to $Z_2$.

The brane description leads to a similar structure. The fivebranes and the $D4$-branes
ending on them combine into a smooth curved fivebrane \WittenEP\ whose form is given by
\eqn\quantumfive{vw=\zeta^2;\qquad v=\zeta e^{-z/\lambda_M}~,}
where 
\eqn\defzz{z=x^6+i x^{11},} 
and $\lambda_M=g_sl_sM=RM$, with $R$ the radius of the M-theory circle, $x^{11}\simeq x^{11}+2\pi R$. We assume that
the IIA string coupling is small but $g_sM$ is large, so that $R=g_sl_s$ is small but the characteristic 
size of the fivebrane \quantumfive, which is governed by $\zeta$, $\lambda_M$, is large (in string units).

Since the position of the fivebranes in $x^6$ does not approach a constant value at large $v, w$, we need 
to impose a UV cutoff on the brane configuration. One way to do that \AharonyMI\ is to define the radial 
coordinate 
\eqn\defuu{u^2=|v|^2+|w|^2=2\zeta^2\cosh{2x^6\over\lambda_M}~,}
and take it to be bounded, $u\le u_\infty$. The curved fivebrane \quantumfive\ must satisfy the boundary
condition  
\eqn\boundcond{\Delta x^6(u_\infty)=L_1~.}
We assume that $L_1<2\pi R_6$, \ie\ the distance between the fivebranes at the cutoff
scale is smaller than the size of the circle. The profile of the brane is schematically
exhibited in figure 7.

\bigskip

\ifig\loc{The quantum ground state of the brane system with $p=0$ is described by the curved fivebrane
\defuu, \boundcond. }
{\epsfxsize4in\epsfbox{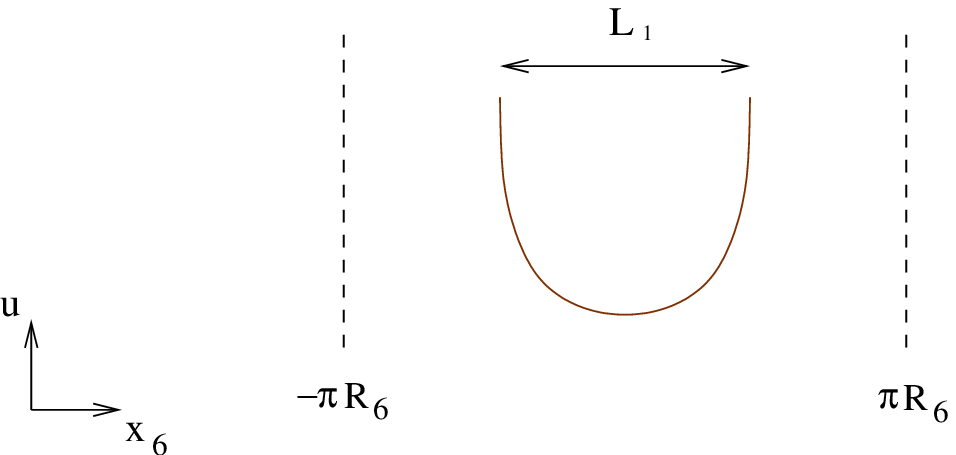}}

\bigskip

Note that the quantum theory is defined by specifying the parameters $M$, $\lambda_M$, 
$u_\infty$, $R_6$ and $L_1$. The dynamical scale $\zeta$ is a derived quantity, 
and can be calculated in terms of these parameters by using \defuu:
\eqn\xiuv{\zeta=u_\infty\exp(-L_1/2\lambda_M)=u_\infty\exp(-1/2\lambda_M^{(4)})~,}
where we defined the four dimensional 't Hooft coupling  in the usual way 
\GiveonSR, $\lambda_M^{(4)}=\lambda_M/L_1$, and assumed that it is small. 
One can think of $\lambda_M^{(4)}$ as the coupling at the UV cutoff scale,
$u_\infty$. The coupling at an arbitrary scale $u$ can be similarly defined by 
replacing $\Delta x^6(u_\infty)=L_1$ by the distance between the two arms of 
the curved fivebrane in figure 7, $\Delta x^6(u)$. For large $u$ it takes the form 
\eqn\largeucurve{{1\over \lambda_M^{(4)}(u)}\simeq 2\ln{u\over\zeta}\simeq 
{1\over \lambda_M^{(4)}(u_\infty)}+ 2\ln{u\over u_\infty}~.}
The preceding discussion is very similar to what happens in gauge theory, where the role of 
$u$ is played by the RG scale, $u_\infty$ is the UV cutoff, and $\lambda_M^{(4)}$ the 't Hooft 
coupling.  The analog of the relation \xiuv\ then gives the QCD scale of the theory, which
we denoted by $\Lambda_1$ in \dynsup; the analog of \largeucurve\ governs the RG flow of
the gauge coupling.

As is well known \WittenEP, the fivebrane \quantumfive\ actually describes a system with $M$ vacua, 
associated with multiplying $\zeta$ by an $M$'th root of unity. These $M$ vacua correspond to
the ones labeled by $r$ in \dynsup. 

\subsec{$0<p<M$}

The gauge theory has in this case three scales:
the dynamically generated scales of the two factors in the gauge group \ggroup, $\Lambda_1$, 
$\Lambda_2$, and the superpotential coupling $h$ (which has units of inverse energy). Due to holomorphy, the moduli space can be studied for any ratio of these scales. A convenient regime
is one in which the gauge coupling of $SU(N_2)$, $g_2$, and Yukawa coupling $h$, are small at the
scale of $SU(N_1)$, $\Lambda_1$. In that case we can first analyze the $SU(N_1)$ dynamics, and then
add the other interactions. 

Since for $p<M$ the $SU(N_1)$ theory has fewer flavors than colors, we can describe the 
supersymmetric vacua in terms of the $2p\times 2p$ meson matrix 
\eqn\mesmat{M_{\alpha\dot\alpha b}^a=A_{\alpha i}^aB_{\dot\alpha b}^i~.}
The superpotential for these fields takes the form 
\eqn\wweff{W_{\rm eff}=W_0+(M-p)\left(\Lambda_1^{3M+p}\over\det M\right)^{1\over M-p}~.}
The F-term constraints of \wweff\ lead to $M$ vacua, which can be thought of as the $M$ vacua 
of the $SU(M)$ pure SYM theory that appears at a generic point in the classical moduli space 
discussed in section 2. 

The mesons \mesmat\ transform in the adjoint ($+$ singlet) representation of $SU(N_2)$. Their $SU(N_2)$ 
dynamics is weakly coupled at low energies. The main effect of this dynamics is to impose the D-term
constraints that, along the moduli space, allow one to diagonalize them (in $a,b$) for all $\alpha$, $\dot\alpha$. 

Thus, the moduli space is labeled by the eigenvalues $M_{\alpha\dot\alpha a}^a$, $a=1,\cdots, p$, 
which satisfy the constraints (that follow from \wweff)
\eqn\constmes{h\det_{\alpha\dot\alpha} M_{\alpha\dot\alpha a}^a=
\epsilon_{M,p}(r, l=0)\sim\left(h^p\Lambda_1^{3M+p}\right)^{1\over M}~,}
\ie\ they lie on the deformed conifold, with deformation parameter $\epsilon_{M,p}(r, l=0)$. 
$r$ is an index that labels the $M$ vacua related by a broken $Z_M$ symmetry, as above. 
The role of the parameter $l$ will become clear shortly.

\bigskip

\ifig\loc{The quantum moduli space of the brane system with $M>p>0$ is described by $p$ $D4$-branes 
wrapping the $x^6$ circle in the vicinity of the curved fivebrane of figure 7. }
{\epsfxsize4in\epsfbox{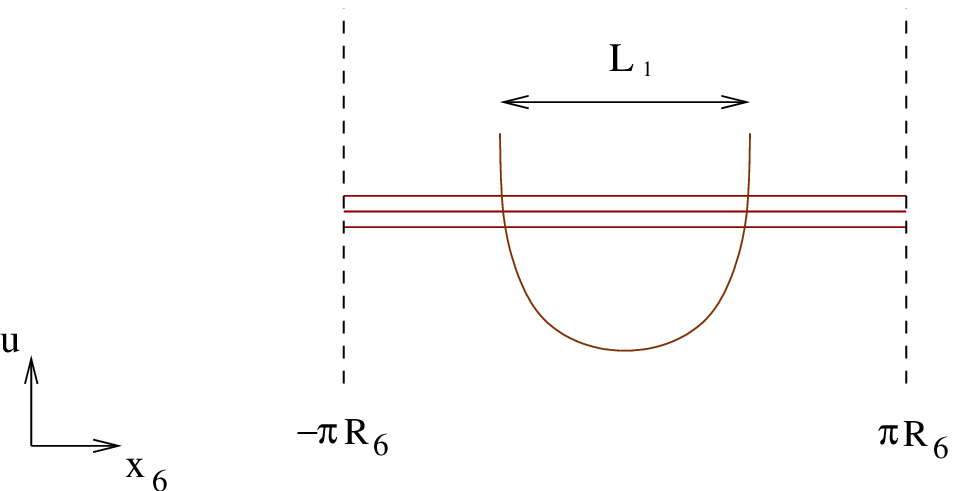}}

\bigskip

To describe the moduli space of vacua in the brane language we need to turn on $g_s$ effects
in the system of figure 2. This involves replacing the $NS5$-branes connected by $M$ $D4$-branes
by the curved fivebrane \quantumfive. The $p$ $D4$-branes in figure 2 then propagate in the vicinity
of this fivebrane (see figure 8). Hence, their moduli space is the deformed conifold (as implied by T-duality).
We conclude that the quantum generalization of the classical moduli space \clasmod\ is
\eqn\quantmod{\MM=\oplus_{r=1}^M {\rm Sym}_p(\CC_{r, l=0})~,}
where $\CC_{r, l=0}$ is the deformed conifold
\eqn\defcon{\det z_{\alpha\dot\alpha}=\epsilon~,}
with deformation parameter $\epsilon=\zeta^2$. We see that the structure of the moduli space agrees with 
that found in gauge theory. 

\subsec{$p=M$}

The gauge theory analysis of \DymarskyXT\ leads in this case to a moduli space of the form 
\eqn\quantmodmm{\MM=\oplus_{l=0}^1\oplus_{r=1}^M  {\rm Sym}_{M(1-l)}(\CC_{r, l})~.}
It is obtained by noting that the $SU(M+p)$ factor in \ggroup\ has 
equal numbers of colors and flavors. Thus, the $SU(N_1)$ dynamics leads at low energies to a $\sigma$-model for the mesons $M$ \mesmat, and baryons $\CA=A^{N_1}$, $\CB=B^{N_1}$. The classical moduli space, which is labeled by $M$, $\CA$, $\CB$, subject to the relation $\det M=\CA\CB$, is deformed in the quantum theory to 
\eqn\defmodspace{\det M-\CA\CB=\Lambda_1^{2N_1}~.}
Adding the effect of the superpotential $W_0$ \www, which is quadratic in the mesons, leads to 
two types of vacua. The mesonic (or $l=0$ in \quantmodmm) vacua have $\CA=\CB=0$ and 
$\det M=\Lambda_1^{2N_1}$. The $SU(N_2)$ D-terms lead then to a moduli space described 
by the eigenvalues of $M$, as in \constmes, \quantmod. The baryonic ($l=1$) vacua are obtained 
by setting the mesons $M=0$; the baryons then satisfy the constraint $\CA\CB=-\Lambda_1^{2N_1}$. 
The low energy theory is pure $N=1$ $SU(M)$ gauge theory, which gives rise to the $M$ isolated 
vacua labeled by $r$ in \quantmodmm. Note that while in the classical theory the baryonic vacuum 
is identical to the origin of the mesonic branch, in the quantum theory the two are distinct, due to the
deformation \defmodspace. The classical result is recovered in the limit $\Lambda_1\to 0$. 

We now turn to the brane description of the vacua \quantmodmm. The mesonic $(l=0)$ branch is
described in the same way as for $p<M$, by the configuration of figure 8 (the quantum version of 
figure 2), with $p=M$. The baryonic vacua are also easy to describe, following the discussion of 
section 2. We saw there that the classical baryonic vacuum of the gauge theory, \susyaone, 
\susyatwo\ is described  by $D4$-branes with non-zero winding (see figure 4). It is natural to expect 
that something similar happens here. 

In more detail, the baryonic vacua are described by the quantum version of a brane configuration in which
$M$ branes connect the $NS$ and $NS'$-branes while winding once around the circle. The classical 
configuration is indistinguishable from that of figure 1 (with $p=M$), which can also be thought of as the
origin of the mesonic branch of figure 2, but quantum mechanically the two are different. While the mesonic
branch is replaced by the configuration of figure 8, a baryonic vacuum gives rise to that of figure 9.

\bigskip

\ifig\loc{The baryonic vacua of the brane system with $p=M$, viewed in the covering space of the $x^6$ circle. Vertical dashed lines are separated by $2\pi R_6$ and are identified on the circle.}
{\epsfxsize4in\epsfbox{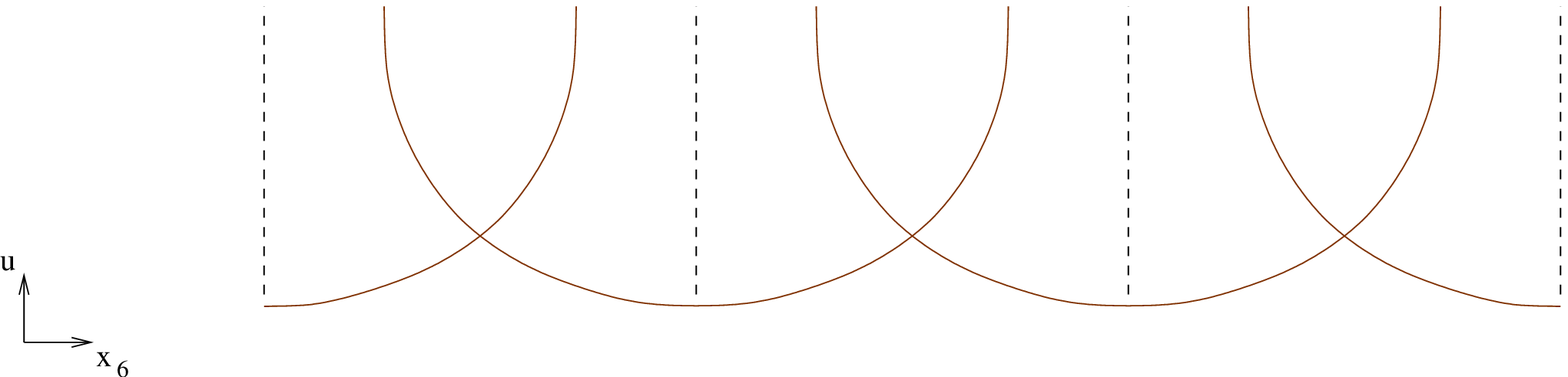}}

\bigskip

In the covering space, it is again described by the profile \quantumfive, but with the boundary conditions 
\boundcond\ replaced by 
\eqn\bcondbar{\Delta x^6(u_\infty)=L_1+2\pi R_6~.}
As in the discussion of section 2, the fact that the curved fivebrane \quantumfive\ winds once
around the circle implies that unlike the mesonic branch of figure 8, here there are no mobile 
$D4$-branes and the vacuum is isolated.  The dynamically generated scale in the baryonic 
vacuum of figure 9 differs from that of the mesonic one (figure 8) as well.  In general, the scale 
is given by (see eq. \xiuv),
\eqn\dynscale{\zeta=u_\infty\exp\left(-\Delta x_6(u_\infty)/2\lambda_M\right)~.}
In the vacua of figures 8,  9 one has 
\eqn\delxsix{\Delta x_6(u_\infty)=L_1+2\pi R_6l~,}
with the winding number $l=0(1)$ in the mesonic (baryonic) branch. 
Plugging \delxsix\ into \dynscale\ we find that 
\eqn\xill{\zeta_l=u_\infty\exp\left(-{\Delta x^6(u_\infty)\over2\lambda_M}\right)=
\zeta_0\exp\left(-{2\pi R_6 l\over 2\lambda_M}\right)=\zeta_0 I^{l\over2M}~,}
with
\eqn\imp{I=\exp\left(-{2\pi R_6\over l_s g_s}\right)~.}
This expression for the scale is the same as that obtained in gauge theory \DymarskyXT.  
We will discuss the general relation in the next subsection. 

An interesting feature of the brane configuration of figure 9 is that there are actually {\it two} different values of the UV cutoff $u_\infty$ for which the two ``arms'' of the curved fivebrane are separated on the $x^6$ circle by the distance $L_1$. One is the value drawn in figure 9, which corresponds to \bcondbar\ and describes a fivebrane that winds once around the circle. The second is obtained by lowering the value of $u_\infty$ until the distance becomes $L_1$ again, this time with no winding. In terms of the dynamically generated scale \xill\ the two values are given by $\zeta_1\exp(L_1+2\pi R_6)/2\lambda_M$ and  
$\zeta_1\exp(L_1/2\lambda_M)$, respectively. For the second (lower) value of the cutoff, for $u<u_\infty$ the brane configuration is identical to the one depicted in figure 7, which describes the vacuum of the theory with $p=0$. Thus, we see that the two are equivalent at long distances; the low energy theory is in both cases $N=1$ pure  $U(M)$ SYM theory.

This infrared equivalence between the $U(2M)\times U(M)$ and $U(M)$ theories can be thought of as a consequence of Seiberg duality. Seiberg duality is usually realized in IIA string theory via motions of fivebranes \ElitzurFH. Here, this motion occurs dynamically, as a function of the RG scale $u$. The situation is under better control than in \ElitzurFH, since the fivebrane configuration of figure 9 remains smooth as $u_\infty$ is decreased. Thus, in this case one does need to rely on unproven conjectures to establish the  equivalence between the baryonic vacua of the theory with $p=M$ and the vacua of the one with $p=0$. 

A few other features of the brane construction are useful to note: 
\item{(1)} While the baryonic $(l=1)$ vacua of the theory with $p=M$ can be identified with those of the $p=0$ one, this equivalence is not true for the mesonic vacua. Indeed, in the configuration of figure 8, the distance on the circle between the two arms of the curved fivebrane is strictly smaller than $L_1$ for all $u$ below the UV cutoff $u_\infty$. There is clearly no corresponding vacuum of the theory with $p=0$.

\item{(2)} In section 2 we discussed what happens when we turn on an FI term for 
$U(1)_b$ in the classical gauge theory. In the quantum theory the situation
is essentially the same. The mesonic branch of moduli space is lifted by the perturbation,
since the mobile fourbranes in figure 8 are no longer mutually BPS with the curved fivebrane,
a rotated version of \quantumfive. The baryonic vacua, which contain no mobile branes, are
still supersymmetric. The curved fivebrane that describes them is the quantum version 
of the classical configuration of figure 4. 

\item{(3)} One could consider {\it increasing} the UV cutoff $u_\infty$ in figure 9, rather than decreasing it, \ie\ flowing up the RG. This relates the vacua of the theory with $p=M$ to those of theories with $p=kM$,  $k>1$. We will discuss such theories next.

\subsec{$p > M$}

For general $M$ and $p$, the gauge theory analysis of \DymarskyXT\ leads to the moduli space
\eqn\genquantmod{\MM=\oplus_{l=0}^k\oplus_{r=1}^M  {\rm Sym}_{p-lM}(\CC_{r, l})~,}
where $k$ is defined in \ptilde, and the deformation parameter of the conifold $\CC_{r,l}$
is given by 
\eqn\epsrl{\epsilon_{M,p}(r, l)=\epsilon_{M,p}(r, l=0)I(M,p)^{l\over M}~,}
with
\eqn\instfactor{I(M,p)=h^{M+2p}\Lambda_1^{3M+p}\Lambda_2^{p-2M}=e^{2\pi i\tau}~.}
The last equality expresses the factor $I(M,p)$ in terms of the D-instanton amplitude
in type IIB string theory. In particular, in string theory this quantity is independent of $M$, $p$.

As before, the index $r$ labels vacua related by the broken $Z_M$ symmetry; the $r$ dependence 
corresponds to picking different $M$'th roots of the identity in \epsrl. 
A natural field theory interpretation of the quantum number $l$ in \genquantmod\ involves 
a series of Seiberg dualities that take $SU(M+p)\times SU(p)$ to $SU(p-(l-1)M)\times SU(p-lM)$.
From the IIB perspective, vacua with given $l$ involve $p-lM$ mobile $D3$-branes propagating 
on the deformed conifold with deformation parameter $\epsilon_{M,p}(r, l)$ \epsrl. 

To describe the vacuum structure \genquantmod\ using the IIA brane construction of figure 1, we 
need to  generalize the discussion of the previous subsections to all $p$.  The parameter $l$ 
labeling different branches of moduli space \genquantmod\ has a clear IIA  interpretation  -- it is the 
winding number of the $D4$-branes connecting the $NS$ and $NS'$-branes.  In a vacuum 
with given $l$, $M$ $D4$-branes stretch from the $NS$-brane to the $NS'$-brane, in the process
winding $l$ times around the circle. This leaves $p-lM$ mobile $D4$-branes wrapping the circle,
which live as before on a deformed conifold. 

The fivebranes with $D4$-branes ending on them are described quantum mechanically 
in terms of a connected curved fivebrane \quantumfive, with the scale parameter $\zeta=\zeta_l$
\xill, \imp. The mobile $D4$-branes live on a deformed conifold \defcon\ with deformation parameter
$\epsilon_{M,p}(r,l)=\zeta_l^2$, which can be written in the form \epsrl, with $I(M,p)$ given by \imp. 
This agrees with the IIB result (the last expression in \instfactor), since one can think of \imp\ as the 
amplitude of a D-instanton obtained by wrapping a Euclidean $D0$-brane around the $x^6$ circle. 
This brane is related by T-duality to the IIB D-instanton whose amplitude is given by \instfactor. 

Note that the deformation parameter goes like $X^l$, with 
\eqn\xxxx{X=I^{1\over M}=\exp\left(-{2\pi R_6\over\lambda_M}\right)~.}
If we choose the 't Hooft coupling at the cutoff scale $\lambda_M^{(4)}$ to be very small, as we have done 
in the discussion around \xiuv, the parameter $X$ is very small as well. Thus, the scales of vacua with 
larger $l$ are strongly suppresed relative to those with smaller $l$. This should be contrasted with the 
situation in the IIB theory where at large 't Hooft coupling (the supergravity regime), the analog of $X$ 
\xxxx\ is very close to one, and one has to consider large values of $l$ to get large suppression. 

We see that the IIA brane description reproduces the structure of the supersymmetric moduli space 
\genquantmod, and the dependence of the deformation parameter \epsrl\ on the branch (\ie\ on $r$
and $l$). One can also
compare the value of the superpotential in the different vacua.\foot{The value of the superpotential is
important for calculating the tension of BPS domain walls between vacua with different values of $r$
in \genquantmod.}  In the field theory, gluino condensation in a low energy $SU(M)$ subgroup of $G$ 
\ggroup\ leads to the superpotential
\eqn\nonzerow{W=ML_1(M,p)^{1\over M}I(M,p)^{l\over M}~,}
where 
\eqn\lone{L_1(M,p)=h^p\Lambda_1^{3M+p}~.}
In the brane language, the superpotential was computed in \WittenEP\ and is given (up to a universal
overall constant) by 
\eqn\iiasup{W\simeq M\zeta_l^2~.}
Substituting the form of $\zeta_l$ \xill\ into \iiasup, we conclude that the two expressions agree
if we take 
\eqn\formlone{\zeta_0^2\simeq L_1(M,p)^{1\over M}~. }
This identification is natural since the right hand side is nothing but $\Lambda^3$, the 
non-perturbative superpotential of the low energy $SU(M)$ gauge theory in the vacuum with $l=0$.

Comments:
\item{(1)} In section 2 we discussed the classical vacuum structure in the presence of an FI 
D-term.  From the IIA brane perspective, it is  clear that the situation in the quantum theory
is similar. If $p$ is not divisible by $M$ (\ie\ if $\tilde p\not=0$ in \ptilde), the vacuum spontaneously
breaks supersymmetry. We will discuss this case further in the next section. For $\tilde p=0$, 
the vacua with $0\le l<k$  again break supersymmetry, while the vacuum with $l=k$, which 
corresponds to the quantum generalization of the configuration of figure 4, does not 
(it is $M$-fold degenerate, as in \genquantmod). 

\item{(2)} As mentioned above, in field theory the vacua \genquantmod\ with $l>0$ can be understood 
in terms of  Seiberg duality. This too has a natural interpretation in the brane construction, as we saw 
in the previous subsection for $p=M$. A vacuum with given $l$ involves $M$ $D4$-branes connecting 
the fivebranes while winding $l$ times around the $x^6$ circle, making a single curved fivebrane of the 
form \quantumfive, with $\zeta=\zeta_l$ \xill.   By decreasing the UV cutoff while keeping the two arms
of the fivebrane at the same distance on the $x^6$ circle  one obtains a vacuum of the theory with 
$p\to p-M$ and $l\to l-1$ (such that the number of mobile $D4$-branes, $p-lM$, remains fixed). 
Looking back at \xill\ we see that 
\eqn\cascade{u_\infty(p-M)=I^{1\over 2M} u_\infty(p)~.}
This is the IIA manifestation of the duality cascade. 

\item{(3)} In gauge theory there are actually two versions of Seiberg duality. The strong version asserts that the electric and magnetic theories of \SeibergPQ\ are equivalent in the infrared at the origin of moduli space and in the absence of deformations of the Lagrangian. In general one or both of these (conformal) theories are strongly coupled, and their equivalence has not been proven to date. The weaker version concerns the infrared equivalence of the two theories in the presence of deformations, and/or along moduli spaces of flat directions. In this case, one can often analyze the long distance behavior of both theories precisely and show their equivalence. Examples of this were studied in the original work of \SeibergPQ\ and many subsequent papers. A discussion in a context closely related to the cascading gauge theory appears in \GiveonEF. In the IIA brane description of Seiberg duality \ElitzurFH, the strong version of Seiberg duality involves exchanging $NS$ and $NS'$-branes connected by $D4$-branes, which involves fivebranes intersecting at a point in the extra dimensions. The statement of duality is that this process is smooth, which is non-trivial and unproven to date. The weak version of the duality involves smooth deformations of the brane system, which obviously do not change the low energy behavior. The discussion above makes it clear that the cascading gauge theory only requires the weak version of the duality. This is why it is manifest in the brane description. It also makes it clear that while the authors of \DymarskyXT\ used Seiberg duality to derive the vacuum structure of the model, one should be able to do this without that assumption, and show that the resulting vacuum structure exhibits the correct duality structure.

\item{(4)} There are many other aspects of the gauge theory that can be studied in the brane description, 
such as domain walls connecting different vacua, QCD strings etc. This description is also useful for
discussing generalizations of the Klebanov-Strassler construction to other cascading gauge theories. 
For example, one can replace the $NS-D4-NS'$ system in figure 2 by a more general one, with or without supersymmetry,
and repeat the discussion of the last two sections.

\newsec{Non-supersymmetric brane configurations}

In the previous section we focused on supersymmetric vacua of the quantum theory. In this section we would 
like to comment on some aspects of the non-supersymmetric dynamics. 

\subsec{Non-supersymmetric vacua with $\xi\not=0$}

In section 2 we discussed the classical theory with non-zero FI parameter for $U(1)_b$.
We saw that the vacuum structure depends on whether $p$ is a multiple of $M$ \pkm. If it is, the lowest
energy state is supersymmetric; it is described by the field configuration \susyaone, \susyatwo\ in the
gauge theory, and by the brane configuration of figure 4 in the IIA language. On the other hand, if 
$\tilde p$ in \ptilde\ does not vanish, the ground state is non-supersymmetric; it is described by  the 
brane configuration of figure 6 and  corresponding field configuration (discussed around \dtermpot).

It is interesting to study the quantum generalization of this brane configuration. An important effect 
that needs to be taken into account in this case is the interaction between the $M-\tilde p$ 
$D4$-branes that wind k times around the circle, and the $\tilde p$ $D4$-branes that wind $k+1$ times.
Since the two stacks of fourbranes are no longer parallel, there is a force between their endpoints on 
the $NS5$-branes. This force is due to an incomplete cancellation between the electrostatic repulsion 
between the endpoints, which can be thought of as (like) charges on the fivebrane, and the attraction 
due to scalar exchange. The former is independent of the angle between the two stacks of $D4$-branes, 
while the latter goes like $\cos\theta$, the angle between the two stacks. 

Thus, the total force is repulsive, and goes like $1-\cos\theta$. This force was discussed in a different
context in \GiveonWP, where this repulsion played an important role in comparing the dynamics of the branes
to that of the corresponding low energy field theory. There, it gave rise to a runaway of certain pseudomoduli;
in our case, the $D4$-branes cannot escape to infinity, since the two fivebranes they are connecting are
stretched in different directions. Thus, the effect of the repulsion is to push them away from each other
by a finite distance. 

This has a natural interpretation in the low energy field theory of the brane system of figure 6. As 
mentioned in section 2, this theory is an $SU(M-\tilde p)\times SU(\tilde p)\times U(1)_b$ gauge
theory coupled to fermions in the bifundamental representation. These fermions are classically massless,
but quantum mechanically are expected to acquire a mass due to chiral symmetry breaking.
The separation of the two stacks of $D4$-branes leads to precisely this effect. The chiral symmetry broken
by the vacuum is part of the $9+1$ dimensional Lorentz group corresponding to rotations in $(45)$ and $(89)$.

One can in principle study the quantum deformations of the configuration of figure 6 in more detail when the
parameters $M$ and $\tilde p$ are in particular regimes. For example, if $g_sM$ is large while $\tilde p$ is of
order one, one can replace the $NS5$-branes connected by $M-\tilde p$ $D4$-branes in figure 6 by a 
curved fivebrane, which looks like a rotated version of \quantumfive, and study the shape of the $\tilde p$ 
probe $D4$-branes which end on this fivebrane and wind $k+1$ times around the circle. If both $g_sM$ and
$g_s\tilde p$ are large, we can replace them by a two center solution and look for the lowest energy 
configuration with the given boundary conditions.  We will leave these calculations to future work. 

The authors of \DymarskyXT\ proposed to use the system with non-zero FI parameter as a possible model of 
early universe cosmology. It is interesting to reexamine this proposal in the regime of validity of the IIA brane construction. 
Consider, for example, the model with $\tilde p=1$, \ie\ $p=kM+1$ (see \ptilde), $k\gg 1$ and $\xi\not=0$. 
For $\xi=0$, the quantum moduli space has multiple branches \genquantmod, most of which are 
unstable for non-zero $\xi$.  In the IIA brane picture, the mobile branes are attracted to the curved fivebrane, 
and are absorbed by it as described in section 2 (figure 5). Even if the FI parameter is not small, \ie\ the  
relative displacement of the $NS5$-branes in figure 3, $\Delta x^7$, is comparable to the distance between the
fivebranes $L_1$, as the process of figure 5 takes place, the angle the curved fivebrane makes with the $x^6$
axis decreases, and thus the attractive potential felt by the mobile $D4$-branes becomes more flat.  

Consider the final step in this process, where all fourbranes but one have been absorbed by the winding
curved fivebrane, which takes the (quantum generalization of the) shape in figure 4, with winding $k$. 
The remaining single mobile fourbrane is subject to a long range attractive potential proportional to 
$1-\cos\theta_k$, where $\theta_k$ is the relative angle between the mobile and bound $D4$-branes,
\eqn\thetak{\tan\theta_k={\Delta x^7\over L_1+2\pi R_6 k}~.}
For large $k$ this angle goes like $1/k$,
\eqn\smallangle{\theta_k\simeq {\Delta x^7\over 2\pi R_6 k}~.}
Since the $M$ bound fourbranes wind $k$ times around the circle, the attractive potential felt by the mobile
fourbrane goes like $V\sim kM(1-\cos\theta_k)\sim M/k$.  Thus, as mentioned above, it becomes more and
more flat as $k$ increases. It would be interesting to see whether it can be made sufficiently flat for inflation to take place. 

The inflationary potential $V$ is due to gravitational attraction between the branes. Thus, it corresponds to a 
D-term potential in the low energy effective description. Therefore, the dynamics studied  here is similar to that
discussed in \refs{\BinetruyXJ, \HalyoPP}, where it was noted that such models have favorable properties in
supergravity (\ie\ at finite $G_N$). 

In this picture, the exit from inflation occurs when the mobile $D4$-brane reaches the vicinity of the curved 
fivebrane. There, processes of the sort depicted in figure 5 transfer the energy of the fourbrane to the fivebrane 
and reheat the universe. The endpoint of the dynamical process is a non-supersymmetric gauge theory, with 
gauge group $SU(M-1)\times U(1)$ and fermions in the adjoint $+$ bifundamental representation. 

It is natural to ask whether the early universe cosmology of the model is likely to lead to the type of initial 
conditions assumed in the above discussion. We will only comment on this issue here, leaving a more detailed
study for future work (see \refs{\AbelCR\CraigKX\FischlerXH-\KutasovKB} for recent discussions of some relevant
issues). At high temperature the system is expected to be in the state with the largest number of massless 
degrees of freedom, which has the lowest free energy. For the moduli space \genquantmod\ this is the branch 
with the largest number of mobile $D4$-branes, \ie\ the one with $l=0$ (figure 8). At zero temperature and 
$\xi\not=0$ 
this is not a true minimum of the energy function, but at high temperature this instability is washed out by thermal 
effects. As the temperature decreases, it becomes less stable, and eventually more and more of the mobile 
$D4$-branes undergo the process of figure 5 and collapse onto the curved fivebrane. Thus, an initial state of
the sort assumed in the discussion of inflation above is not particularly unnatural in the early universe evolution
of this system. 

\subsec{Adding ${\bar D}$-branes to KS}

The authors of \KachruGS\ proposed that adding anti $D3$-branes to the type IIB brane system of 
\KlebanovHB\ leads to the appearance of metastable states in which the antibranes expand into 
an $NS5$-brane which can only annihilate via quantum tunneling. Much about these states remains 
mysterious. In the IIB gravity regime, the approximations employed in \KachruGS\ to establish their 
existence are not obviously reliable. If  these states do exist, 
there is the question whether they should be thought of as metastable states in the Klebanov-Strassler 
gauge theory, or as states in a bigger theory that also contains the supersymmetric KS states.  

In this subsection we will study these issues in the IIA description. Our conclusions will not 
be directly applicable to the IIB regime, or to the gauge theory, since the different regimes  are related 
by large continuous deformations, which may well change the energy landscape. Nevertheless, it seems 
useful to address these questions in any regime where they can be analyzed reliably. 

We start with the brane system studied in the previous sections, with 
\eqn\pbar{p=kM-\bar p;\qquad 0<\bar p<M~.}
We saw that this system has a rich moduli space of vacua \genquantmod, labeled among other things
by the number of mobile $D4$-branes $p-lM$, $l=0,\cdots, k-1$. Since this number never vanishes, all 
the vacua \genquantmod\ belong in this case to mesonic branches. 

Following \KachruGS, we  start with the vacuum with $l=k-1$, which has $M-\bar p$ mobile $D4$-branes,
and add $\bar p$ pairs of $D4$ and $\bar{D4}$-branes wrapping the circle. The brane configuration now 
contains $M$ $D4$-branes and $\bar p$ $\bar{D4}$-branes, and there are two possible things that can 
happen to it:
\item{(1)} The antibranes can annihilate with some of the branes. This takes us back to the mesonic
supersymmetric vacuum with $M-\bar p$ mobile $D4$-branes. 
\item{(2)} The $M$ $D4$-branes can combine with the curved fivebrane \quantumfive, and increase its
winding from $k-1$ to $k$. This describes the baryonic vacuum of the theory with $p=kM$, but now
we also have $\bar p$ $\bar{D4}$-branes propagating in the vicinity of the curved fivebrane. 

\noindent
The second possibility gives rise to the metastable state of \KachruGS.
The $\bar{D4}$-branes, which wrap the $x^6$ circle,  are T-dual to the $\bar{D3}$-branes discussed in 
\KachruGS. Placing the $\bar{D3}$-branes at the tip of the conifold corresponds in the IIA language to 
placing the $\bar{D4}$-branes at $u=0$ (see figure 10). In the IIB description it was argued in 
\KachruGS\ that the antibranes expand into an $NS5$-brane carrying $\bar{D3}$-brane charge. The 
IIA analog of this phenomenon is the following. 

\bigskip

\ifig\loc{The baryonic branch of the brane system with $p=kM$,  with $\bar p$ $\bar{D4}$-branes wrapping
the circle ($k=1$ in the figure). }
{\epsfxsize4in\epsfbox{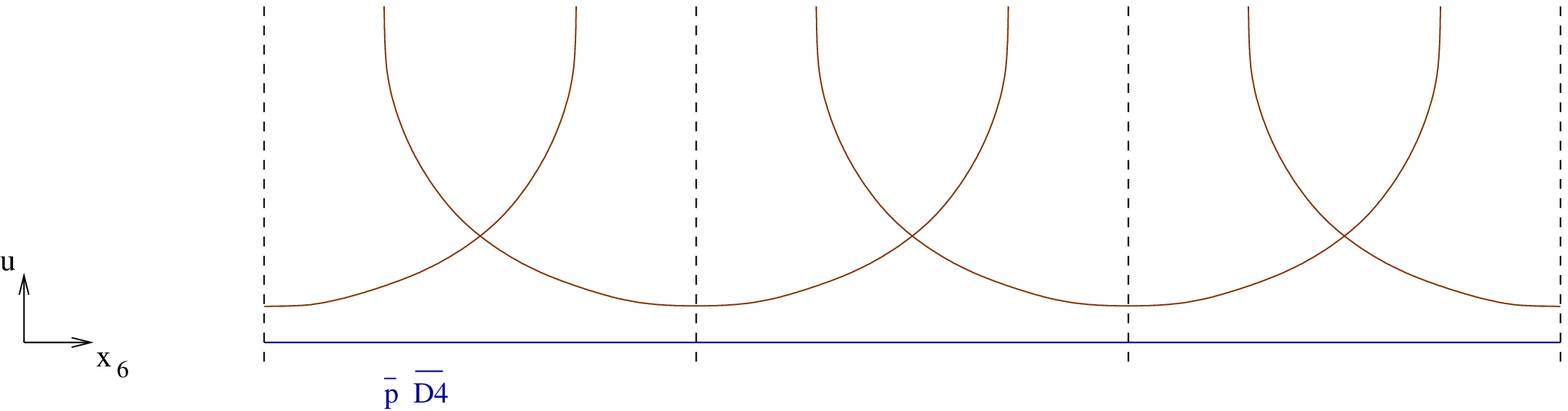}}

\bigskip

While the configuration of figure 10 is stationary, it is not stable. The $\bar{D4}$-branes are attracted to the
curved fivebrane (which carries fourbrane charge), and if one displaces them infinitesimally from $u=0$,
will start moving towards the fivebrane. Consider, for example, the case $\bar p=1$. The lowest energy
configuration of the single $\bar{D4}$-brane is qualitatively described by the configuration of figure 11.   
It can be determined in the probe approximation; we will not describe the details here. 
The $D4$-brane flux carried by the bottom of the fivebrane in figure 11 is $M-\bar p$; the location of the 
$\bar{D4}$-brane is determined by balancing the geometric and electrostatic forces acting on it. 

\bigskip

\ifig\loc{The configuration of figure 10 is unstable to decay to that depicted here. }
{\epsfxsize4in\epsfbox{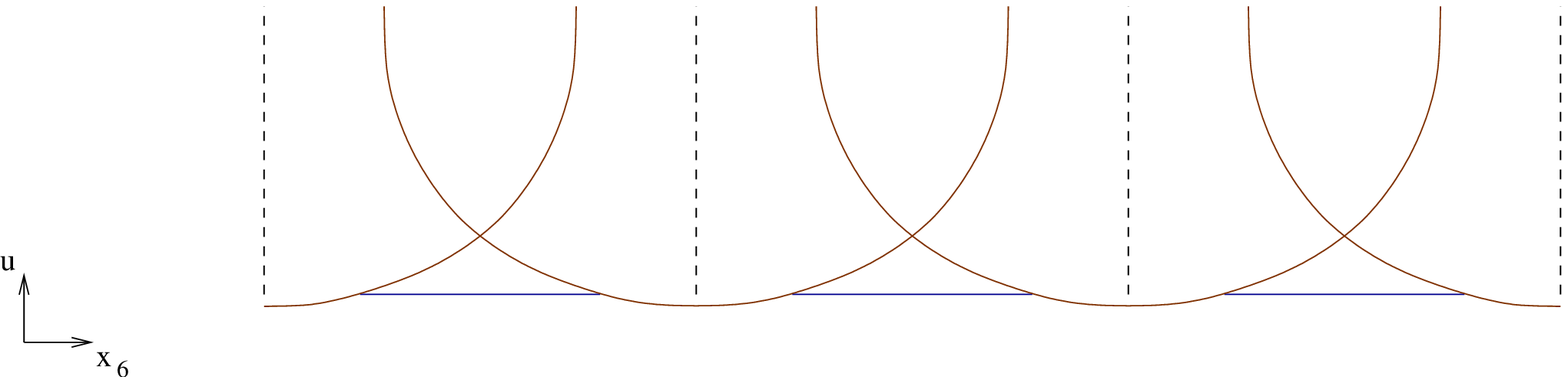}}

\bigskip

Since the $\bar{D4}$-brane is displaced from the origin of the $\IR^4$ labeled by $(v,w)$ \vvww, the configuration
of figure 11 breaks the $U(1)$ symmetry of the curved fivebrane \quantumfive, which acts as (opposite) 
rotations in $v$, $w$. The interpretation of this symmetry in the gauge theory was discussed in \AharonyMI. 
Its breaking gives rise to a Nambu-Goldstone boson, which corresponds to slow motions of the $\bar{D4}$-brane
on the circle of fixed $u(x^6)$ corresponding to its shape. 

If the number of $\bar D4$-branes, $\bar p$, is larger than one,\foot{But much lower than $M$, so that we can
neglect their backreaction on the shape of the fivebrane.} each of the $\bar{D4}$-branes can be analyzed as 
above. Since the different $\bar{D4}$-branes repel each other \GiveonSR, they arrange themselves into a discretized
tube connecting the two sides of the curved fivebrane. This is the IIA manifestation of the $NS5$-brane carrying
$\bar p$ units of $\bar{D}$-brane charge of \KachruGS. The configuration of figure 11 is locally stable, but 
can decay via tunneling to the supersymmetric mesonic branch with $M-\bar p$ mobile $D4$-brane described
above. 

The dynamics described by the brane configuration of figure 11 in various energy regimes can be
understood by starting at small $u$ (low energy) and studying the configuration as we increase $u$. 
For $u$ below the position of the antibranes, the brane configuration is identical to that of figure 7, 
\ie\ it corresponds to pure $N=1$ SYM with gauge group $SU(M-\bar p)$. As we increase $u$, we 
get to the position of the $\bar{D4}$-branes (blue line in figure 11). Above the corresponding energy, 
we can think of the brane system as describing the quantum vacuum of the brane system of figure 12.

\bigskip

\ifig\loc{The low energy description of the metastable vacuum of figure 11 consists (classically) 
of $M-\bar p$ $D4$-branes (red) and $\bar p$ $\bar{D4}$-branes (blue) stretched between 
the $NS5$-branes. }
{\epsfxsize3.5in\epsfbox{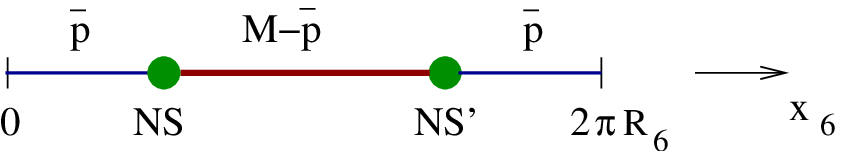}}

\bigskip
The effective gauge theory in this regime is an $SU(M-\bar p)\times SU(\bar p)\times U(1)$ gauge theory
with fermions in the adjoint and bifundamental representation of the gauge group. The bifundamental
fermions are classically massless (figure 12), but quantum mechanically they acquire a mass via chiral 
symmetry  breaking. This is the field theory analog of the fact that the antibranes are located at a finite 
value of $u$ in figure 11. Continuing to larger $u$, the brane configuration approaches the baryonic 
vacuum of the theory with $p=lM$, with $l$ increasing up to $k$ at the UV cutoff scale $(u=u_\infty)$. 
The $\bar{D4}$-brane gives rise to a localized perturbation of the curved fivebrane \quantumfive.

Interestingly, the effective field theory that describes the metastable SUSY breaking state of \KachruGS,
which corresponds to the brane configuration of figure 12, is the same as the low energy theory of the
supersymmetric system with non-zero FI parameter $\xi$ (discussed after eq. \ptilde), with $\bar p$ here 
playing the role of $\tilde p$ there. From the brane perspective, this is very natural -- the two are related 
by a continuous deformation. 
Indeed, starting with the configuration of figure 6, one can move the fivebranes towards each other,
such that the winding number $k$ decreases. It is clear from the figure that no states go to zero 
mass in the process; thus, the low energy theory is unchanged by this deformation. Eventually, the 
winding number of the $M-\tilde p$ $D4$-branes vanishes. If we go once more around
the circle, these branes reverse their orientation, and we end up with a configuration similar to that
of figure 12, with the two $NS5$-branes displaced relative to each other in $x^7$. However, it is clear
from figure 12 and the analysis of \gkunpub\ that this displacement also does not change the low
energy spectrum and dynamics. 

Thus, we see that the brane systems of figures 6, 11 correspond  to different UV completions of the 
$SU(M-\bar p)\times SU(\bar p)\times U(1)$ gauge theory described above. In particular, in figure 6 
(and 12) supersymmetry is broken in the ground state, while in figure 11 the same low energy theory 
arises as an effective infrared theory in a metastable ground state. 

An interesting and widely discussed question is whether the metastable state of \KachruGS\ is a state in the cascading gauge theory (see \eg\
\refs{\BenaXK\DymarskyPM\BlabackNF-\MassaiJN}). In the IIA regime the answer appears to be negative for the following reason. The gauge theory provides a low energy description of the brane system of figure 1, or its quantum version discussed in section 3. While one can arrange the parameters of the model such that the metastable state of figure 1 has a small energy density, the height of the barrier for the tunneling to the supersymmetric state is determined by the energy (density) of $\bar p$ $D/\bar D$ pairs wrapping the circle. For $\bar p=1$ this energy is (in string units) $E\sim R_6/g_s$. Using \couplbar\ one can write it as $E\sim 1/g^2 k$, where $g$ is the four dimensional gauge coupling of the low energy theory. Thus, for finite $g, k,$ the height of the barrier between the supersymmetric and non-supersymmetric vacua is finite in string units, and hence the tunneling between the two goes to zero in the gauge theory limit. This should be contrasted with the situation in brane constructions of metastable vacua that are visible in the gauge theory, such as that of \GiveonEW, where all energy scales, including the height of the barrier, can be taken to be small. Thus, we conclude that while the configuration of figure 11 is metastable in the full string theory, it is stable in the low energy theory. It corresponds to a different superselection sector of the theory on the branes from the supersymmetric vacua.

\newsec{$N=2$ cascade}

$N=2$ supersymmetric gauge theories are known to exhibit cascading behavior similar to that found for $N=1$ in \KlebanovHB\ (see \eg\ \refs{\PolchinskiMX\BeniniIR-\DasguptaSW}). At first sight this is puzzling, since $N=2$ supersymmetric QCD does not exhibit Seiberg duality. As we saw above, the type IIA description provides a useful guide for studying the classical and quantum vacuum structure of cascading gauge theories. In this section we will use it to shed light on the $N=2$ duality cascade. 

\bigskip

\ifig\loc{The IIA brane configuration that realizes the $N=2$ supersymmetric cascading gauge theory.}
{\epsfxsize3.3in\epsfbox{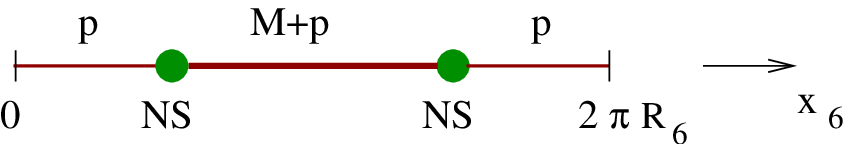}}

\bigskip

The brane configuration corresponding to the gauge theory we are interested in is a close analog of that of figure 1, and is depicted in figure 13. The different branes are oriented as in section 2 (see \braneor); the fact that the $NS5$-branes are parallel implies that this configuration preserves eight supercharges, or $N=2$ supersymmetry in the $3+1$ dimensions $(0123)$. The low energy theory is in this case an $N=2$ SYM theory with the gauge group\foot{We will include in the gauge group the $U(1)$ factors, which were omitted in \ggroup.} and matter content \ggroup\ -- \transfab. The superpotential \www\ is now absent and is replaced by the standard $N=2$ superpotential that couples the adjoints in the vector multiplet of $G$ \ggroup\ to the bifundamentals \transfab. If one breaks $N=2$ SUSY by giving a mass to the adjoints (which corresponds in the brane picture to a relative rotation of the two fivebranes in $(v,w)$) one can recover \www\ by integrating them out.   

In the rest of this section we will repeat the discussion of sections 2, 3 for the $N=2$ supersymmetric case, and describe the classical and quantum supersymmetric vacuum structure of the brane system of figure 13. It should be clear from the $N=1$ analysis above, and from the study of many other systems reviewed in \GiveonSR, that the results apply to (and can be stated in terms of) the low energy $N=2$ SQCD. The brane picture merely provides a useful language for describing the vacuum structure. 

\subsec{Classical moduli space} 

The basic fact that governs the classical moduli space of the brane configuration of figure 13 is that fourbranes stretched between the two fivebranes (``fractional branes'') are free to move along the fivebranes, in the $v$ plane \vvww, while fourbranes that wrap the whole circle (``regular branes'') are free to move in the whole transverse $\IR^5$ labeled by $(45789)$. An example of a branch of the classical moduli space is  the Coulomb branch for the two gauge groups, which corresponds in figure 13 to displacing the $M+p$ coincident $D4$-branes to arbitrary positions $v_i$, $i=1,\cdots, M+p$, and the $p$ $D4$-branes connecting the fivebranes on the other side of the circle to $\tilde v_a$, $a=1,\cdots, p$. At  a generic point in this moduli space (with all $v$, $\tilde v$ distinct) the gauge group is broken to $U(1)^{M+2p}$. When one of the $v$'s and one of the $\tilde v$'s coincide, a bifundamental hypermultiplet goes to zero mass and a new branch of moduli space opens up. In the brane language it corresponds to the two fractional branes connecting into a regular brane, which can move off the fivebranes into the aforementioned $\IR^5$. 

\bigskip

\ifig\loc{The brane description of $\MM_n$, a component of the classical moduli space, has $n$ fourbranes wrapping the circle moving in the transverse $\IR^5$ (in general away from the fivebranes), and $M+p-n$ resp. $p-n$ fourbranes connecting the fivebranes and distributed in the $v$ plane.}
{\epsfxsize3.3in\epsfbox{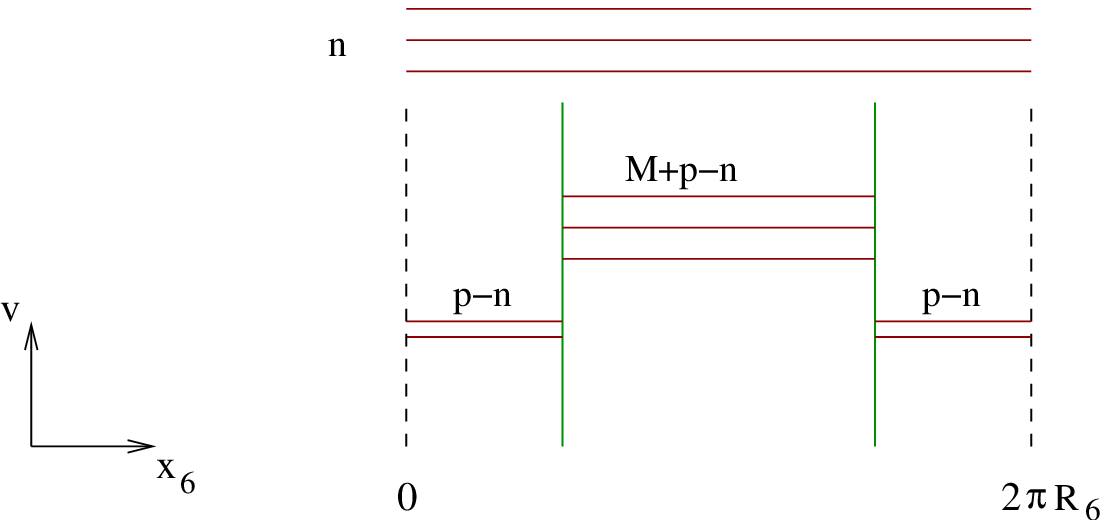}}

\bigskip

The full classical moduli space is a direct sum of spaces $\MM_n$, $n=0,1,2,\cdots, p$, which are described in the brane language by the configuration of figure 14. As is clear from the picture, at a generic point in the moduli space the low energy theory includes $N=4$ SYM with gauge group $U(1)^n$, and pure $N=2$ SYM with gauge group $U(1)^{M+2p-2n}$. The different $\MM_n$ intersect on subspaces where some charged hypermultiplets go to zero mass, and have other singular points at which charged vector multiplets become massless and enhance the gauge group. 

As in the $N=1$ case, the FI coupling $\xi$ for $U(1)_b$ corresponds in the brane picture of figure 13 to a relative displacement of the two $NS5$-branes in $x^7$. In general, this leads to non-supersymmetric vacua of the sort discussed in section 2 (around figure 3), while for $p=kM$ with integer $k$ one finds supersymmetric vacua of the sort discussed around figure 4. These vacua involve fourbranes connecting fivebranes while winding ($k$ times) around the $x^6$ circle. 

In the $N=2$ case one can also displace the fivebranes in the $(89)$ plane. This corresponds in the gauge theory to turning on a linear superpotential $W=\lambda{\rm Tr}\Phi$ for the chiral superfield in the $U(1)_b$ vector multiplet. It has a similar effect on the vacuum structure to that of the FI term. In fact, $(\xi,\lambda)$ transform as a triplet under the $SU(2)_R$ symmetry of the $N=2$ SYM theory, which corresponds in the brane language to the rotation symmetry  $SO(3)_{789}$.

\subsec{Quantum moduli space}

Going from classical to quantum gauge theory corresponds in the IIA brane system to turning on a finite string coupling $g_s$. When one does that, the system of two $NS$-branes connected by $N$ $D4$-branes becomes a single connected $NS5$-brane carrying $D4$-brane charge \WittenMT.\foot{In \WittenMT, $g_s$ was taken to be large. In this limit the bulk spacetime becomes eleven dimensional, and the fivebrane in question becomes an $M5$-brane. As discussed in \AharonyMI, one can alternatively consider the limit $g_s\ll1$, $g_sN\gg1$, in which the right description is in terms of an $NS5$-brane in weakly coupled type IIA string theory.} For example, the Coulomb branch discussed above, which corresponds to figure 14 with $n=0$, is described by a curved fivebrane which looks asymptotically (at large $v$) like a pair of curved $NS5$-branes with the profile $z\sim \pm\lambda_M\ln v$ (see the discussion around eq. \defzz\ for the notation), connected by $M+p$ respectivelly $p$ tubes.  The precise form of the curved fivebrane is described in \WittenMT. 

In the $N=1$ supersymmetric case we saw (in section 3) that the quantum vacuum structure is richer than the classical one. The basic reason for that is that configurations which are identical in the classical limit become distinct at finite $g_s$. In particular, the classical configuration of $M$ $D4$-branes connecting the fivebranes with $M$ additional fourbranes wrapping the circle and intersecting the fivebranes, can be viewed as the classical limit  of either the Higgs  branch (figure 8) or the baryonic branch (figure 9). We expect the same to happen in the $N=2$ case. 

Consider, for example,\foot{It is easy to generalize the discussion to other branches of the moduli space.} the branch of moduli space with $n=p$ in figure 14. In this branch, the theory generically reduces at low energies to a direct product of $N=2$ SYM with gauge group $SU(M)$ along its Coulomb branch, and $p$ copies of $U(1)$ $N=4$ SYM. The configuration of figure 14 describes the classical moduli space; quantum mechanically, the fourbranes connecting the two fivebranes become finite tubes. Together with the $NS5$-branes they make the curved fivebrane \WittenMT
\eqn\curvefive{t^2+B(v)t+1=0,}
where $t=\exp(-z/R)$ and $B(v)=v^M+u_2v^{M-2}+\cdots +u_M$. As in the $N=1$ case, we can introduce a UV cutoff by taking $|v|$ to be bounded, $|v|\le v_\infty$, and demand that the distance between the two arms of the curved fivebrane at the cutoff scale is equal to some fixed length $L_1<2\pi R_6$, ``the distance between the fivebranes''.  If the moduli $u_2,\cdots, u_M$ are small relative to the cutoff scale $v_\infty$, one has 
\eqn\distarms{L_1\simeq 2\lambda_M\ln (v_\infty/\zeta), }
with $\zeta$ a scale that was set to one before. 

Following the discussion of the $N=1$ case, one can obtain additional branches of the quantum moduli space by taking $lM$ of the $p$ mobile $D4$-branes to coincide with the $M$ $D4$-branes stretched between the fivebranes, and consider the quantum configuration corresponding to $M$ fourbranes connecting the two $NS$-branes while winding $l$ times around the circle, together with $p-lM$ mobile $D4$-branes in the bulk of the $\IR^5$.  For $p$ of the form \ptilde, the maximal value of $l$ is $l_{\rm max}=k$, and if $\tilde p=0$, one has in that case a close analog of the baryonic branch of the $N=1$ supersymmetric theory of section 3. The low energy theory in this branch is pure $N=2$ SYM with gauge group $SU(M)$, and the moduli space is its Coulomb branch. The curved fivebrane is again described by \curvefive, but now the distance between the two arms at the UV cutoff scale, which enters \distarms,  is $L_1+2\pi k R_6$, as in the $N=1$ discussion. Hence the fivebrane winds $k$ times around the $x^6$ circle.

An important difference with respect to the $N=1$ discussion is that for $N=2$, every time the curved fivebrane winds around the circle it intersects itself at $2M$ points.\foot{To find these points one needs to calculate the intersections of the curve \curvefive\ with another copy of this curve, in which $t\to It$ (see \imp\ for the definition of $I$).} This self intersection is very similar to the one discussed in appendix A. As there, each intersection point supports a $U(1)$ vector multiplet and a massless charged hypermultiplet. 

The rest of the discussion is similar to the $N=1$ case. The fivebrane \curvefive\ that winds $l$ times around the circle describes a particular branch of the moduli space of the theory corresponding to figure 13 with $p=kM$. By decreasing the value of the UV cutoff $v_\infty$ one can also view it as a vacuum of the theory with $p=(k-1)M, (k-2)M$, etc, together with $2M, 4M,\cdots$ decoupled sectors consisting of a vector multiplet and a charged hypermultiplet . If we neglect these decoupled sectors, we conclude that the theories with $p=lM$ with different values of $l$ share part of their moduli space of  vacua. Of course, there are some branches of the moduli space that are different as well. That was already the case in the $N=1$ case \DymarskyXT, but for $N=2$ there are more branches of moduli space, and naturally more of them are different in theories with different values of $l$. 

To understand the origin of the $N=2$ duality cascade from the point of view of the low energy gauge theory, consider the simplest case $p=M$. The vacuum structure of the resulting $U(2M)\times U(M)$ gauge theory can be analyzed by studying first the limit where the $U(M)$ gauge coupling is very small. Then we have a $U(2M)$ $N=2$ SQCD with $N_f=2M$ flavors. As we review in appendix A, this theory has a baryonic branch, whose root is described at low energies by a $U(1)^{2M}$ gauge theory with $2M$ hypermultiplets charged under the different $U(1)$ factors \ArgyresEH, see eq. (A.1). These fields are all singlets under the $SU(N_f)$ global symmetry. Thus, gauging $U(M)$ does not influence them, and the full low energy theory at the root of the baryonic branch is a direct product of the above abelian sector and the Coulomb branch of pure $U(M)$ $N=2$ SYM. This picture is in complete agreement with the brane description above. The baryonic branch of the moduli space is described by a curved fivebrane \curvefive\ that winds once around the circle. The abelian factors live at the $2M$ self intersections of this curve, while the small $v$ shape of the fivebrane describes the Coulomb branch of the low energy $U(M)$ pure SYM. Clearly, one can iterate this procedure to describe the vacuum structure of theories with larger $p$, as was done in \DymarskyXT\ for the $N=1$ case.

To summarize, if one neglects the abelian sectors, one finds that the $U(lM)\times U((l-1)M)$ gauge theories at the root of their baryonic branches are all equivalent, and flow in the IR to pure $U(M)$ $N=2$ SYM; this equivalence is manifest in the brane description.  This is the origin of the cascading behavior seen in the IIB description in  \refs{\PolchinskiMX\BeniniIR-\DasguptaSW}. The cascading geometries in these papers appear to describe the dual of the curved fivebrane, whereas the abelian factors that distinguish theories with different values of $l$ presumably correspond to singletons, that live at the boundary of the space.

As mentioned above, the full quantum moduli space of the $N=2$ gauge theory with general $p$ is quite intricate. For example, starting with the classical moduli space $\CM_n$ of figure 14, we can take $l_1(M+p-n)$ of the mobile $D4$-branes and attach them to the $M+p-n$ $D4$-branes stretched between the fivebranes, making them wind $l_1$ times around the circle; similarly we can attach $l_2(p-n)$ of the remaining mobile $D4$-branes to the $p-n$ stretched $D4$-branes in figure 14, and make them wind $l_2$ times around the circle. This gives new branches of moduli space labeled by $(l_1, l_2, n)$, which satisfy
\eqn\newbranches{l_1(M+p-n)+l_2(p-n)\le n~.}
The discussion of this section can be generalized to these vacua as well.

\newsec{Discussion}

The main conclusion of this work is that the IIA brane description provides a useful qualitative and quantitative guide to the dynamics of cascading gauge theories with various amounts of supersymmetry. In particular, we saw that for the $N=1$ cascading theory of \KlebanovHB, the classical and quantum moduli spaces of supersymmetric vacua agree, including the dependence of the dynamically generated scale \epsrl\ on the parameters labeling different  branches of moduli space (which is given in the brane picture by \xill, \imp). The brane picture makes it clear that the cascade utilizes a weak form of Seiberg duality, which involves deformed SQCD, and can be proven regardless of whether the stronger version of the duality holds.

We also saw that the brane picture provides a useful guide to the non-supersymmetric dynamics of the theory.  In particular, we discussed the stable non-supersymmetric vacuum obtained for non-zero FI parameter $\xi$ and generic $p$, $M$, and the dynamics as one approaches it from vacua on the classical pseudo moduli space. It would be interesting to find the IIB geometry corresponding to the stable non-supersymmetric vacuum of figure 6, which describes the brane system with $\tilde p\not=0$. It is natural to expect that when $g_s\tilde p$, $g_s M$ are large, this geometry should be smooth.

We saw (in section 4.2) that the metastable state described in IIB language in \KachruGS\ 
has a IIA analog. The fact that this state exists in the regime of parameter states where the IIA description
is reliable supports the construction of \KachruGS. In the IIA regime this state is clearly metastable,
and decays to the same supersymmetric state as in the proposal of \KachruGS. An interesting open
question is whether this state exists also in the gauge theory. From the IIA point of view this appears
to be unlikely. To get it we added a $D4/\bar{D4}$ pair to the theory with $p=kM-\bar p$. This seems to 
lead to a system with more degrees of freedom than the original $SU(p)\times SU(M+p)\times U(1)$
gauge theory. This is reflected in the fact that the height of the barrier between the non-supersymmetric and supersymmetric vacua goes to infinity in the gauge theory limit ($m_s\to\infty$ with the gauge coupling held fixed). We also noted that the low energy dynamics of the metastable  state is closely related to that of the non-supersymmetric state at non-zero $\xi$. As we saw, this is very natural from the brane 
description.  

In section 5 we generalized the discussion to systems with $N=2$ supersymmetry. The type IIA description clarifies why they exhibit cascading behavior despite the fact that Seiberg duality is not a symmetry of such theories. This is due to the fact that while the full theory does not exhibit Seiberg duality, certain vacua do. Thus, some of the vacua of the $N=2$ theory with gauge group $U(M+p)\times U(p)$ are 
shared by theories with $p\to p-M, p-2M,\cdots$. Even in these vacua the equivalence is not complete -- theories with higher $p$ differ from those with lower one by a decoupled sector with an abelian gauge group coupled to charged hypermultiplets.

\bigskip

\noindent{\bf Acknowledgements}: We thank O. Aharony, K. Dasgupta, A. Giveon, J. Marsano, N.  Mekareeya,  and J. Seo for discussions.  AW thanks the EFI at the University of Chicago for hospitality  while part of this work was completed.  The work of DK is supported in part by DOE grant DE-FG02-90ER40560 and by the BSF -- American-Israel Bi-National Science Foundation. The work of AW is supported by the McGill University Schulich Fellowship.

\appendix{A}{Aspects of the IIA description of $N=2$ SQCD}

$N=2$ SQCD with gauge group $U(N_c)$ and $N_f$ hypermultiplets in the fundamental representation of the gauge group can be described by the brane configuration of figure 15. 

\bigskip

\ifig\loc{The brane description of $N=2$ SQCD with $N_c$ colors and $N_f$ flavors.}
{\epsfxsize3.3in\epsfbox{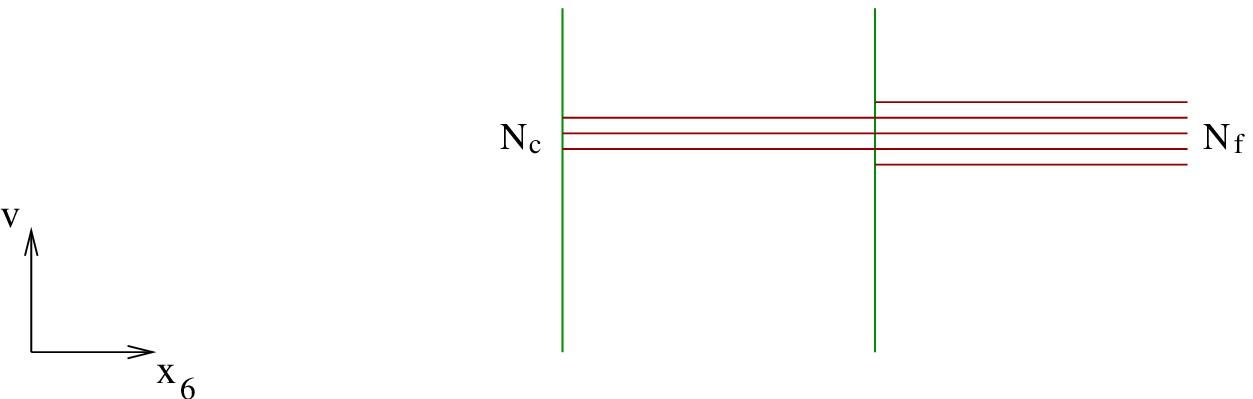}}

\bigskip

\noindent
It consists of two $NS$-branes (see \braneor\ for the orientations of the  branes) connected by $N_c$ (``color'') $D4$-branes, which give rise to $N=2$ SYM with gauge group $U(N_c)$. $N_f$ (``flavor'') $D4$-branes attached to one of the fivebranes give hypermultiplets in the fundamental representation of the gauge group. In order to study the full moduli space of vacua of the theory one needs to terminate the $N_f$ flavor branes on $D6$-branes, but since this is not going to be important for our purposes, we will keep them semi-infinite. 

The classical and quantum vacuum structure of $N=2$ SQCD was analyzed in \ArgyresEH.\foot{These authors studied the case of $SU(N_c)$ gauge group, but the theory with gauged baryon number is closely related.} Our main interest is going to be in the parameter range $N_c<N_f<2N_c$, and in the baryonic branch, in which the gauge symmetry is in general completely broken. At the origin of this branch the classical theory has an unbroken $U(N_c)$ gauge symmetry, but quantum effects are large. The authors of \ArgyresEH\ showed that in the quantum theory, the origin of the baryonic branch has an alternative weakly coupled description with gauge group 
\eqn\ttlldd{U(\tilde N_c)\times U(1)^{N_c-\tilde N_c};\qquad \tilde N_c=N_f-N_c~.}
The matter consists of $N_f$ hypermultiplets in the fundamental of $U(\tilde N_c)$ which are not charged under the $U(1)$'s, and $N_c-\tilde N_c$ hypermultiplets $e_i$ which are singlets of $U(\tilde N_c)$ and charged under the $U(1)$'s (the latter can be normalized such that $e_i$ has charge $-\delta_{ij}$ under the $j$'th $U(1)$). 

\bigskip

\ifig\loc{The brane description of $N=2$ SQCD at finite $\xi$.}
{\epsfxsize3.3in\epsfbox{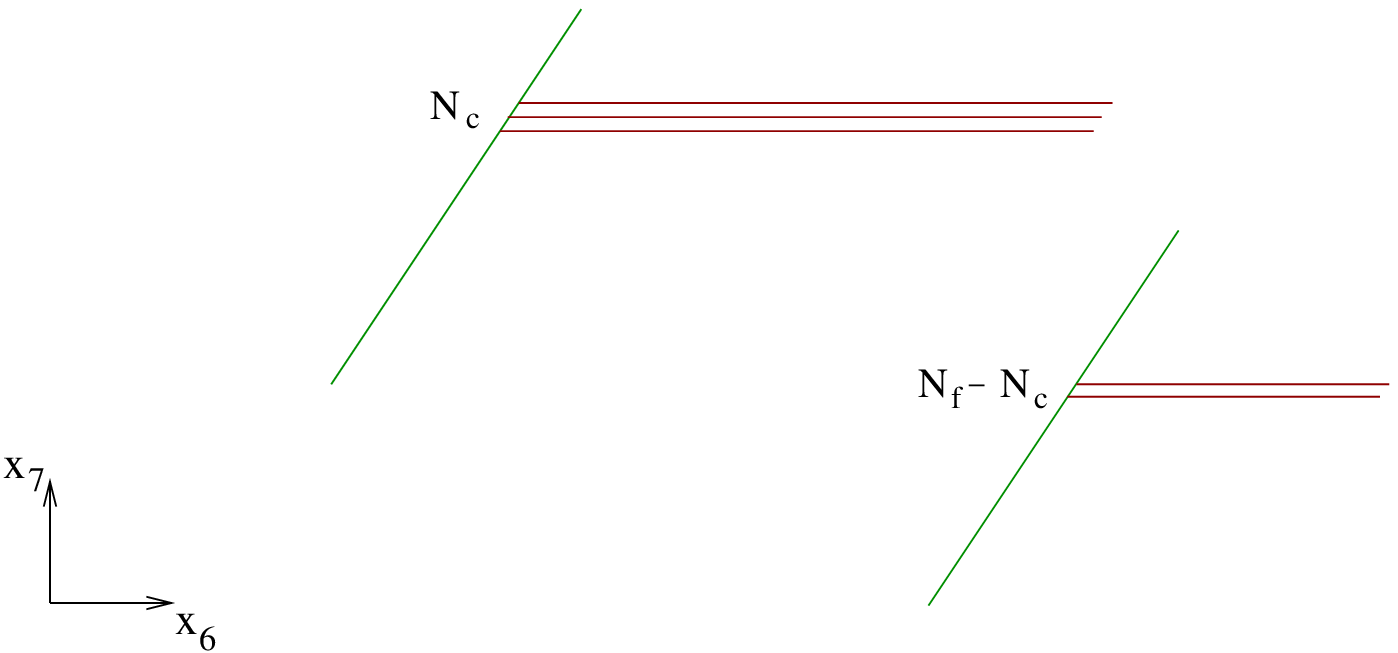}}

\bigskip

From the brane perspective, this can be understood as follows. As discussed in the text, one can take the theory into the baryonic branch by turning on the FI parameter $\xi$, which corresponds in the brane description to a relative displacement of the $NS$-branes in the $x^7$ direction. For finite $\xi$ the $U(N_c)$ gauge symmetry is broken and the brane system splits into two disconnected components (see figure 16). As $\xi\to 0$ the $U(N_c)$ gauge symmetry is restored, and quantum effects become important. Thus, we have to replace the brane system of figure 16 by its finite $g_s$ analog \refs{\WittenEP,\WittenMT}. The two fivebranes in figure 16 take the forms
\eqn\formnc{v^{N_c}=t~,\qquad v^{\tilde N_c}=\zeta^{\tilde N_c} t~,}
respectively. Here we used the freedom of choosing the origin in $x^6$, $x^{11}$ to set the coefficient of $t$ to one for one of the two fivebranes. The constant $\zeta$ can be determined by imposing the boundary conditions that at $|v|=|v_\infty|$ the two fivebranes are separated by the distance $L$, 
\eqn\formzeta{|\zeta|^{\tilde N_c}={e^{L/ R}\over |v_\infty|^{N_c-\tilde N_c}}~.}
Viewed in the $(x_6, |v|)$ plane, the fivebranes take the form depicted in figure 17. 

\bigskip

\ifig\loc{The origin of the baryonic branch of $N=2$ SQCD in the quantum theory. The dashed line corresponds to the UV cutoff $|v|=|v_\infty|$.}
{\epsfxsize2.2in\epsfbox{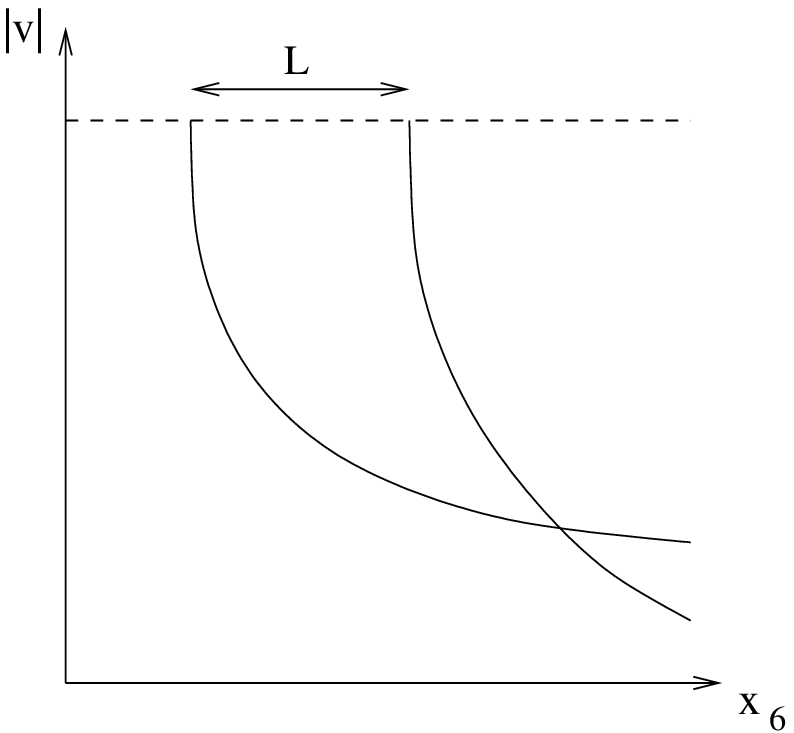}}

\bigskip

\noindent
The fact that the two fivebranes approach each other as $|v|$ decreases reflects the growth of the gauge coupling in the infrared. At some point the fivebranes intersect and cross, and for smaller $|v|$ (\ie\ low energy), their ordering in $x^6$ is reversed. As we further lower $|v|$, the distance between the fivebranes increases, reflecting the infrared freedom of the low energy effective theory. To see what that theory is we need to take the classical limit of the resulting brane configuration, which is depicted in figure 18. 

 \bigskip

\ifig\loc{The classical limit of the small $v$ limit of the brane configuration of figure 17.}
{\epsfxsize3.3in\epsfbox{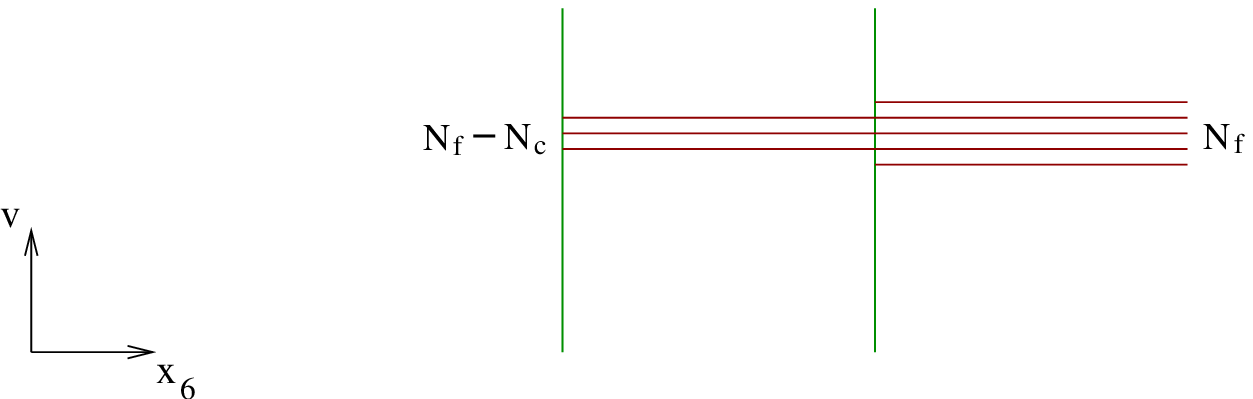}}

\bigskip
 
\noindent
It is a $U(\tilde N_C)$ $N=2$ SQCD with $N_f$ flavors, which is indeed not asymptotically free (and is thus weakly coupled in the IR). This theory is very similar to that found in \ArgyresEH, \ttlldd, but it is missing the $U(1)$ factors in the gauge group and the charged hypermultiplets $e_i$. 

It is clear from figure 17 that these must come from the fivebrane intersection. While it seems from the figure that the two component fivebranes intersect at a single point, in fact there are $N_c-\tilde N_c$ intersection points (at finite $t$), which can be obtained by imposing both equations in \formnc. This gives 
\eqn\intersect{v^{N_c-\tilde N_c}=\zeta^{-\tilde N_c}~,}
which has $N_c-\tilde N_c$ solutions lying on a circle of fixed $|v|$. Comparing to \ttlldd, it is natural to conjecture that each intersection supports a $U(1)$ vector multiplet and a charged hypermultilet. It should be possible to show this directly in string theory, but we will not attempt to do this here. 

Note that here we are interpreting the radial direction transverse to the $D4$-branes, $|v|$, as parametrizing energy, with small (large) $v$ corresponding to low (high) energies. It may seem peculiar from this point of view that some of the massless degrees of freedom, namely the $U(1)$ factors in \ttlldd\ live at finite $v$, \intersect.  This phenomenon is actually familiar from the study of brane systems in string theory. The non-abelian degrees of freedom associated with such systems (say, the $SU(N_c)$ part of the gauge group) typically live in the near-horizon region of the branes, while the $U(1)$ factors are localized in the interface between the near and far regions.

To summarize, the brane system of figure 15 provides a simple way to understand the dual description of the root of the baryonic branch of $N=2$ SQCD \ttlldd. This description is in the spirit of \ElitzurFH; the non-abelian factor in the dual gauge group arises from brane exchange (which happens here as a function of RG scale), and the $U(1)$ factors and charged hypermultiplets live at self-intersections of the quantum fivebrane.\foot{The brane description also makes it clear that the dual description of the root of the baryonic branch \ttlldd\ is related to the microscopic $N=2$ SQCD in a simpler way than the magnetic Seiberg dual theory \SeibergPQ\ is related to the microscopic electric theory in the $N=1$ supersymmetric case. In particular, while in the former case one can derive the dual (or effective low energy) description from the microscopic one, in the latter no such derivation is known.} Turning on a FI term in the microscopic theory corresponds in the low energy description to a FI term for the overall $U(1)$ in $U(\tilde N_c)$ and all the $U(1)$ factors in \ttlldd, which Higgses the gauge group and gives masses to the $e_i$. In the brane description this corresponds to separating the two component fivebranes in figure 17 in $x^7$, so they no longer intersect, and all degrees of freedom associated with the intersections become massive.

The authors of \ArgyresEH\ also discussed what happens to the theory when one breaks $N=2$ supersymmetry down to $N=1$ by giving a mass to the adjoint chiral superfield in the $(S)U(N_c)$ vector multiplet. In the brane description this corresponds to rotating one of the $NS$-branes in figure 15 from the $v$ to the $w$ plane. Since the fivebranes are no longer parallel, the curves in figure 17 do not intersect in the extra dimensions. This is the brane reflection of the fact that in this case the charged chiral hypermultiplets $e_i$ get a non-zero vev,  Higgs the $U(1)$ gauge group, and lift to non-zero mass all states associated with the intersections.

\listrefs
\end